\documentclass{aa}  
\usepackage{graphicx}
\usepackage{txfonts}
\usepackage{xcolor}
\usepackage{hyperref}
\usepackage[normalem]{ulem}
\hypersetup{colorlinks=true,urlcolor=blue,linkcolor=blue,citecolor=blue}

\def \MSUN{\rm M_{\odot}}

\def \MFIVEC{M_{\rm 500c}}
\newcommand{\mytilde}{\raise.19ex\hbox{$\scriptstyle\sim$}}

\begin{document} 

   \title{Radio relics in massive galaxy cluster mergers \\in the TNG-Cluster simulation}
   \titlerunning{Radio relics in TNG-Cluster}

   \author{W. Lee
          \inst{1,2}\thanks{\email{wonki.lee@yonsei.ac.kr, mkjee@yonsei.ac.kr}},
          A. Pillepich \inst{3},
          J. ZuHone \inst{2},
          D. Nelson \inst{4},
          M. J. Jee \inst{1,5}, 
          D. Nagai \inst{6},
          K. Finner \inst{7}
          }
   \authorrunning{W. Lee et al.}
   
   \institute{Yonsei University, Department of Astronomy, Seoul, Republic of Korea
    \and
    Harvard-Smithsonian Center for Astrophysics, 60 Garden St., Cambridge, MA 02138, USA
    \and
    Max-Planck-Institut f{\"u}r Astronomie, K{\"o}nigstuhl 17, D-69117 Heidelberg, Germany
    \and
    Zentrum f{\"u}r Astronomie der Universit{\"a}t Heidelberg, ITA, Albert Ueberle Str. 2, D-69120 Heidelberg, Germany
    \and
    Department of Physics, University of California, Davis, One Shields Avenue, Davis, CA 95616, USA
    \and
    Department of Physics, Yale University, New Haven, CT 06520, USA
    \and 
    IPAC, California Institute of Technology, 1200 E California Blvd., Pasadena, CA 91125, USA
             }

  \date{}

  \abstract
   {Radio relics are diffuse synchrotron sources in the outskirts of merging galaxy clusters energized by the merger shocks. In this paper, we present an overview of the radio relics in massive cluster mergers identified in the new TNG-Cluster simulation.
   This is a suite of magnetohydrodynamical cosmological zoom-in simulations of 352 massive galaxy clusters with $\MFIVEC = 10^{14.0-15.3}~\MSUN$ sampled from a 1 Gpc-sized cosmological box. The simulations were performed using the moving-mesh code AREPO with the galaxy formation model and high numerical resolution consistent with the TNG300 run of the IllustrisTNG series. We post-processed the shock properties obtained from the on-the-fly shock finder to estimate the diffuse radio emission generated by cosmological shockwaves for a total of $\mytilde300$ radio relics at redshift $z=0-1$.
   TNG-Cluster returned a variety of radio relics with diverse morphologies, encompassing classical examples of double radio relics, single relics, and ``inverted" radio relics that are convex to the cluster center. Moreover, the simulated radio relics reproduced both the abundance and statistical relations of observed relics. We find that extremely large radio relics ($>$ 2 Mpc) are predominantly produced in massive cluster mergers with $\MFIVEC\gtrsim8\times10^{14}~\MSUN$. This underscores the significance of simulating massive mergers to study giant radio relics similar to those found in observations. We released a library of radio relics from the TNG-Cluster simulation, which will serve as a crucial reference for upcoming next-generation surveys.}

   \maketitle
%
%-------------------------------------------------------------------

\section{Introduction}

Collisions between clusters of galaxies have been among the most energetic processes since the Big Bang. 
Cluster mergers inject a large amount of energy, on the order of $\mytilde10^{64}\rm~erg$, to the cluster environment and play a fundamental role in the evolution of galaxy clusters \citep[e.g.,][]{2001ApJ...561..621R,2002ASSL..272....1S}. 
The collision between two or more clusters causes the distribution of the collisionless components, such as the dark halo and galaxies, as well as the collisional intracluster medium (ICM), to elongate along the collision axis \citep[e.g.,][]{1997ApJS..109..307R}. 
With a nonzero pericenter separation, the collision impacts the angular momentum and triggers a rotational motion that facilitates the mixing of the ICM \citep[][]{2001ApJ...561..621R,2006ApJ...650..102A,2010ApJ...717..908Z}. 
Gas compression due to a supersonic collision velocity ($\mytilde3,000\rm~km~s^{-1}$) generates merger shocks, which further dissipate the energy in the cluster outskirts \citep[e.g.,][]{2000ApJ...542..608M,2003ApJ...593..599R,2007PhR...443....1M}.

A fraction of the energy dissipated by merger shocks is believed to generate a relativistic plasma via the diffusive shock acceleration (DSA) process \citep{1983RPPh...46..973D,1987PhR...154....1B}. 
For $\mytilde\rm\mu G$ intracluster magnetic fields, the relativistic electrons -- that is, cosmic ray electrons (CRe) with a Lorentz factor of $\gamma_{\rm L}\sim10^{4}$ radiate via synchrotron emission and form megaparsec-sized diffuse radio emissions called radio relics. 

Radio relics are highly polarized ($30-60\%$), elongated radio features that are most often observed in cluster outskirts \citep[][]{2012A&ARv..20...54F,2019SSRv..215...16V}.
It is now well accepted that the formation of radio relics is closely related to merger-driven shocks \citep[e.g.,][]{1998A&A...332..395E,2008A&A...486..347G}. 
Cluster mergers launch a pair of shockwaves propagating in opposite directions along the collision axis \citep[e.g.,][]{2011MNRAS.418..230V,2018ApJ...857...26H}, which can explain the spatial distribution of the radio relics with respect to the cluster X-ray and weak-lensing (WL) mass distributions \citep[e.g.,][]{2015ApJ...802...46J,2021ApJ...918...72F,2022ApJ...925...68C}.
Moreover, compression by shock waves aligns the turbulent intracluster magnetic field with the shock front \citep[][]{1980MNRAS.193..439L,2010Sci...330..347V}. 
The shock acceleration of CRe creates radio emission with a power-law spectrum \citep[e.g.,][]{1987PhR...154....1B,2020A&A...636A..30R}.
Independent detections of shock fronts from X-ray observations across radio relics further support the merger shock-radio relic connection \citep[e.g.,][]{2013MNRAS.433..812O,2015A&A...582A..87A,2015MNRAS.449.1486S,2016ApJ...818..204V,2016MNRAS.460L..84B,2016MNRAS.463.1534B,2019ApJ...873...64D,2019MNRAS.489.3905S,2023A&A...670A.156C}. 

This formation scenario of radio relics is supported by both cosmological and idealized simulations of merging clusters.
Simulations have shown that radio emission enabled by shock acceleration resembles the general properties of radio relics, including the luminosity function, scaling relations, and the Mach number discrepancy between X-ray and radio observations \citep[e.g.,][]{2008MNRAS.391.1511H,2009MNRAS.393.1073B,2011ApJ...735...96S,2012MNRAS.421.1868V,2013ApJ...765...21S,2015ApJ...812...49H,2015MNRAS.451.2198V,2017MNRAS.470..240N,2018ApJ...857...26H,2019MNRAS.490.3987W,2021MNRAS.506..396W}. 
Moreover, simulations are successful tools for modeling plasma acceleration efficiency. 
In fact, observations report that cluster merger shocks may be too weak ($\mathcal{M}<3$) to explain the observed radio brightness under the DSA process \citep[e.g.,][]{2020A&A...634A..64B}. 
Two scenarios have been proposed to resolve this inefficiency problem: 1) a reacceleration of fossil CRe, where an old CRe population with relatively low energy (i.e., fossil CRe) boosts the acceleration efficiency in weak shocks \citep[$\mathcal{M}<3$, e.g.,][]{2011ApJ...734...18K}; or 2) a multiple-shock (MS) scenario, where successive passages of merger shocks enhance the radio emissivity \citep[e.g.,][]{2021JKAS...54..103K,2022MNRAS.509.1160I}.  
These models are implemented in numerical models and can therefore be evaluated through comparison with observations \citep[e.g.,][]{2013MNRAS.435.1061P,2016ApJ...823...13K,2020ApJ...894...60L,2021MNRAS.500..795D,2023Galax..11...45V,2023A&A...669A..50V,2023MNRAS.519..548B}.

However, recent discoveries of radio relics from state-of-the-art radio observations with unprecedented sensitivity and resolution reveal several discrepancies compared to existing simulations. 
First, previous simulations have not been able to reproduce extremely large radio relics.
Many observed radio relics exhibit the largest linear sizes (LLS) exceeding $\gtrsim2\rm~Mpc$ \citep[e.g.,][]{2001A&A...376..803G,2010Sci...330..347V,2013ApJ...769..101V,2014ApJ...785....1B,2016ApJ...818..204V,2021A&A...656A.154H,2022A&A...657A..56K,2022A&A...659A.146D}, with the largest radio relic ever detected reaching $\mytilde3.5\rm~Mpc$ \citep[e.g.,][]{2021A&A...656A.154H}.
Second, previous simulations lack the diverse morphology observed in real radio relics.
Observations often reveal radio relics that are misaligned with the hypothesized collision axis \citep[e.g.,][]{2018ApJ...859...44H,2023A&A...675A..51D} or present inhomogeneous radio emission \citep[e.g.,][]{2015MNRAS.449.1486S,2018MNRAS.478.2218H}. These features suggest an inhomogeneous acceleration efficiency due to the presence of fossil CRe \citep[e.g.,][]{2014ApJ...785....1B,2015MNRAS.449.1486S,2017NatAs...1E...5V,2019ApJ...874..112S}. 
In addition, recent observations have identified filamentary substructures, as seen in Abell 2256 \citep{2022ApJ...927...80R} and Abell 3667 \citep{2022A&A...659A.146D}, for example, and in ``inverted'' radio relics. These inverted relics exhibit an arc shape, but are convex toward the center of the cluster, as seen in the Coma cluster \citep[e.g.,][]{2022ApJ...933..218B}, SPT-CL J2023-5535 \citep[e.g.,][]{2020ApJ...900..127H}, Abell 1697 \citep[e.g.,][]{2021A&A...651A.115V}, the Ant Cluster \citep[e.g.,][]{2021ApJ...914L..29B}, and Abell 3266 \citep[e.g.,][]{2022MNRAS.515.1871R}.
Finally, the rarity of the radio relics is still an open question. Recently, \citet{2023A&A...680A..31J} claimed that only two systems out of 46 clusters at $z>0.5$ host radio relics, whereas cosmological simulations suggest a significantly larger fraction \citep[$\mytilde50\%$,][]{2012MNRAS.420.2006N}. 

Simulating these radio relics with a high fidelity in the cosmological context is a challenging task. 
While bright radio relics are predominantly reported in the most massive clusters with $M_{\rm 200c}>10^{15}\rm~\MSUN$ \citep[e.g.,][]{2015ApJ...802...46J,2021A&A...656A.154H,2023A&A...675A..51D}, these extreme populations are rare in simulations, and limited by the simulation volume \citep[e.g.,][]{2023A&A...673A.131L}.
Given that only a small ($\lesssim10\%$) fraction of clusters are expected to host radio relics \citep[e.g.,][]{2015A&A...579A..92K,2022MNRAS.517.1299Z,2023A&A...680A..31J}, a substantial number of clusters must be explored to enable statistically robust analyses and to sample rare morphologies. 
In addition to the large simulation volume required, the resolution must be high enough to resolve the merger shocks in rarified cluster outskirts. 
A robust model of radiative physics and AGN/stellar feedback of the cluster galaxies is also required, as they can modify the ICM profile and the merger shocks that propagate in the ICM \citep[e.g.,][]{2015JKAS...48..155K,2019MNRAS.488.5259Z}.

In this paper, we use TNG-Cluster, a new suite of cosmological magnetohydrodynamics zoom-in simulations of 352 massive clusters, to understand the formation of radio relics in the massive cluster mergers. 
TNG-Cluster adopts the same galaxy formation model and the numerical resolution of the highest resolution run of the TNG300 volume of the IllustrisTNG suite. As we demonstrate, TNG-Cluster provides the largest sample of radio relics ever simulated, including full magnetohydrodynamics and realistic models for galaxy formation and evolution. 
Based on this archive of massive clusters, we study whether a simple shock acceleration model can explain observed giant radio relics, their morphological diversity, and their low occurrence rate, particularly at high redshift. 

This work is part of an introductory series of studies of TNG-Cluster, which aims to demonstrate the scientific potential of the simulation suite in various contexts.
Whereas here we focus on the abundance and properties of radio relics from massive cluster mergers, \citet{2023arXiv231106338N} introduces the simulation project and showcases the general ICM properties of the simulated clusters. 
In \textcolor{blue}{Pillepich et al. (in prep.)}, we explore the global and spatially resolved morphologies of the ICM, including pressure waves and feedback-driven bubbles and perturbations.
\citet{2023arXiv231106339A} provides an atlas of the velocity fields of the cluster gas, including line-of-sight velocities, velocity dispersions, and velocity structure functions.
In \citet{2023arXiv231106334T}, we forecast the velocity dispersion of Perseus-like cluster cores that will be observed with the XRISM mission \citep{2020arXiv200304962X}. A first look at the cool-core versus non-cool-core populations in TNG-Cluster and their connection to cluster properties is quantified by \citet{2023arXiv231106333L}, while \citet{2023arXiv231106337R} studies the circumgalactic medium of cluster satellites and their observable signatures.

This paper is organized as follows.
We describe the TNG-Cluster simulations and the models used to calculate radio emissivity in \textsection\ref{sec:method}. 
An overview of the identified cluster mergers and radio relics is presented in \textsection\ref{sec:result}, while their physical properties and scaling relations are quantified in \textsection\ref{sec: statistics}. 
We discuss the origins of the diverse properties of radio relics in \textsection\ref{sec:discussion} before the summary in \textsection\ref{sec:summary}. 
Appendix \ref{sec: app-res} shows the convergence tests with numerical resolutions, while Appendix \ref{sec: app-obstab} summarizes the properties of observed radio relics that we compare to simulated ones. 

The TNG-Cluster simulation adopts the Planck2015 $\Lambda$CDM cosmology \citep{2016A&A...594A..13P}, namely $h=0.6774$, $\Omega_{\rm m}=0.3089$, $\Omega_{\rm \Lambda}=0.6911$, $\Omega_{\rm b}=0.0486$, $\sigma_{8}=0.8159$, and $n_{\rm s}=0.9667$. 
$R_{\Delta\rm c}$ describes the radius where the average density becomes $\Delta$ times the critical density of the universe, and $M_{\Delta\rm c}$ is the total mass within $R_{\Delta\rm c}$.

%%%%%%%%%%%%%%%%%%%%%%%%%%%%%%%%%%%%%%%%%%%%%%%%%%

\section{Methods}
\label{sec:method}

\subsection{The TNG-Cluster simulation suite}
\label{sec:tngc}

TNG-Cluster\footnote{\url{https://www.tng-project.org/cluster/}} is a spin-off project of the IllustrisTNG series \citep[TNG hereafter: ][]{2018MNRAS.475..648P,2018MNRAS.475..676S,2018MNRAS.480.5113M,2018MNRAS.475..624N,2018MNRAS.477.1206N,2019MNRAS.490.3196P,nelson2019a,2019MNRAS.490.3234N}.
In particular,  magnetohydrodynamics cosmological zoom-in simulations of 352 massive galaxy clusters sampled from a 1 Gpc-size cosmological box in a $\Lambda$CDM scenario.

Specifically, TNG-Cluster focuses on the time evolution of galaxy clusters at the high-mass end, including all 92 halos with $M_{\rm 200c}>10^{15}\rm~\MSUN$ in the parent box at $z=0$. 
For the mass range of $M_{\rm 200c}=10^{14.3-15}\rm~\MSUN$, zoomed-in clusters were randomly selected to make a flat halo mass function when combined with TNG300, which is the largest volume simulation in the IllustrisTNG project. All details about the simulation, sample, and galaxy formation model can be found in \citet{2023arXiv231106338N} -- here we briefly discuss relevant features.

The simulations are performed using the moving-mesh code \texttt{AREPO} \citep{2010MNRAS.401..791S} and
use the same galaxy formation physics as TNG, including thermal and kinetic AGN feedback modes \citep{2017MNRAS.465.3291W} in addition to radiative cooling, star formation, galactic winds, and metal enrichment \citep{2018MNRAS.473.4077P}. The TNG model has been shown to agree well with the observed nature of galaxy clusters in a wide range of contexts \citep[e.g.,][]{2018MNRAS.474.2073V,2018MNRAS.481.1809B,2018MNRAS.481.1950L,2021MNRAS.501.1300S}. In fact, the TNG model has also been shown to return realistic cluster galaxies in terms of stellar and gas content for both brightest cluster galaxies \citep[e.g.,][]{2018MNRAS.475..648P, 2020A&A...641A..60P, 2023MNRAS.521..800M} and satellites \citep[e.g.,][]{2019MNRAS.483.5334S, 2021MNRAS.506.4760D}, providing support for the implemented feedback models. 

Importantly, for the scientific applications of this paper, TNG-Cluster follows the evolution and amplification of magnetic fields by solving the ideal continuum magneto-hydrodynamics (MHD) equations \citep{2011MNRAS.418.1392P}. 
From an initial homogeneous seed of $10^{-14}$ comoving Gauss, magnetic fields are amplified by flux conservation in gravitational collapse, turbulence, and shear flows, resulting in $\mytilde0.1-10\mu G$-scale magnetic fields in the $z\sim0$ ICM \citep{2023arXiv231106338N}.

The mass resolution of the TNG-Cluster is identical to the highest resolution run of TNG300. Dark matter and baryons are resolved with a mass resolution of \mbox{$m_{\rm DM}=6.1\times10^{7}~\MSUN$} and \mbox{$m_{\rm baryon}=1.2\times10^{7}~\MSUN$}, respectively. 
The gravitational softening length of collisionless particles (DM and stars) is $\mytilde1.4\rm~kpc$ at $z=0$, whereas gas cells use an adaptive comoving softening length with a minimum value of $\mytilde0.4\rm~kpc$. The smallest gas cell is $\mytilde70$ pc in size, as cells are smaller at progressively higher densities. 
In general, we find that the merger shockwaves residing at $R_{\rm 500c}$ are resolved with cells smaller than $\mytilde10\rm~kpc$, which are hence capable of tracking shock properties (Appendix \ref{sec: app-res}) but insufficient to fully resolve their filamentary substructures \citep[e.g.,][]{2022ApJ...927...80R,2022A&A...659A.146D}.

Because of the zoom-in nature of the runs, simulation elements from low-resolution regions might contaminate the high-resolution zoom volumes. Throughout this paper, we only use gas within the high-resolution regions. In fact, for the average cluster, the mass fraction of low-resolution DM particles drops below the $10^{-6}$ level at about 3 virial radii, so numerical contamination is not an issue \citep[see][for details]{2023arXiv231106338N}.

Finally, halos and galaxies in TNG-Cluster are identified as in all other TNG simulations, via a friends-of-friends \citep[FoF,][]{1985ApJ...292..371D} and the \texttt{SUBFIND} \citep{2001MNRAS.328..726S} algorithm. The latter locates and characterizes gravitationally bound subhalos. Galaxies are subhalos containing non-zero stellar mass. Subhalos are tracked in time using the \texttt{SubLink} merger trees \citep{2015MNRAS.449...49R}. Snapshots and halo and galaxy catalogs are available at 100 points in time, as in the other TNG simulations \citep[see][for more details]{nelson2019a}: these are separated in time by about 150 Myr at low redshift. However, among these 100, the so-called full snapshots, which include Mach number and magnetic field properties needed for this paper, are sampled more sparsely (one per about 1 Gyr).

\subsection{Identification of merging clusters}
\label{sec:merger_def}

Throughout this paper, we exclusively focus on the 352 systems, which are the primary zoom-in targets of TNG-Cluster, namely, the most massive halos in the zoom-in regions in the snapshots at redshift $z\leq 1$.

We identify merging systems using the \texttt{SUBFIND} and \texttt{SubLink} halo and tree catalogs, respectively. For all \texttt{SUBFIND} objects (hereafter subhalos) within the high-resolution region, we refer to any subhalo that has undergone a first pericenter passage with respect to the main cluster as a collider or merging subcluster or merging subgroup hereafter. 

We compute the gravitationally bound total mass of the collider as follows. 
As a subhalo falls into the main cluster potential, its mass decreases and introduces ambiguity. 
We use the maximum subhalo mass before the first pericenter passage to correctly identify the mass that triggered the collision.
The collider mass typically peaks when the halo separation is around $\mytilde2R_{\rm 200c}$, with the center of each object being defined as the position of the most gravitationally bound resolution element. 
Thus, we define the merger mass ratio as the ratio between the main cluster and the collider mass ($M_{\rm main}/M_{\rm sub}$) derived from the snapshot, at which the subcluster mass is at its maximum.
In the following, our study is limited to merging systems with a subhalo mass greater than $10^{13}\,\MSUN$.

The properties of the merger are determined from the orbit of the collider by measuring the time evolution of its separation from the main cluster.
Because the time step of the simulation output is too sparse to precisely sample the closest passage, the pericenter separation and the time of the pericenter passage are derived by fitting a quadratic function to the five data points closest to the pericenter passage.
We compute the collision velocity using the time derivative of the halo separation, while halo centers are based on the locations of the potential minima.

When a subhalo possesses a dense dark core, the halo finder algorithm incorrectly misidentifies it as the main halo.
This confusion can arise during the disruption of a dense group, which can be detected as a sudden discontinuity in the trajectory of the main cluster.
Similar switching errors can occur for a major merger when the two merging halos have comparable masses. In such cases, we monitor the continuity of the halos in terms of their positions and velocities to identify potential switching errors, and apply corrections to avoid these issues.

As a cluster merger hosts two or more mass peaks with a comparable mass, we redefine the center of the system using the center of mass. 
We start from the center in the halo catalog, which defines the center using the position of the most gravitationally bound particle. Then, we gather the dark matter particles within $R_{\rm 500c}$ and compute the center of the mass for this particle group. 
We iterate the calculation until the consecutive center shift becomes $\lesssim10\rm~kpc$. 
Hereafter, we adopt this new cluster center as the center of the projection maps.

\subsection{Radio emissivity in post-processing}
\label{sec: emissivity}

TNG-Cluster employs an on-the-fly shock finder algorithm \citep{2015MNRAS.446.3992S} that can successfully identify shockwaves in cosmological galaxy simulations \citep{2016MNRAS.461.4441S}. Given the characterized shock properties, we estimate the radio emission by post-processing the simulated merging clusters in a spatially resolved manner. In the following, we describe the implemented methodology. 

Firstly, the shock finder identifies shock zones where the gas cells satisfy the following conditions: (i) a converging gas flow is in place ($\nabla\cdot  v<0$), (ii) temperature and density gradients are aligned ($\nabla T\cdot  \nabla \rho >0$), and (iii) a discontinuity is greater than $\mathcal{M}>1.3$ \citep[e.g.,][]{2003ApJ...593..599R,2008ApJ...689.1063S,2014ApJ...785..133H}.
The shock surface cell with the maximum gas flow convergence within the shock zones is then identified.
The shock normal, namely the direction of shock propagation, is determined with the temperature gradient $\nabla T$ and is measured with second-order accuracy \citep{2016MNRAS.455.1134P}. 
The upstream and downstream regions are recorded using the first cells outside the shock zones on either side along the shock normal, which are used to compute the shock properties. 
The shock strength $\mathcal{M}$ is estimated from the temperature jump between the upstream and downstream cells, in accordance with the Rankine–Hugoniot jump condition. 
The shock dissipation rate ($E_{\rm diss}$) is derived with
\begin{equation}
    E_{\rm diss}=\frac{1}{2} \rho_{1} (\mathcal{M}c_{\rm s,~1})^3 A \delta(\mathcal{M}),
\end{equation}
where $\rho_1$ is the upstream density, $c_{\rm s,~1}$ is the upstream sound velocity, $A$ is the shock surface area, and $\delta(\mathcal{M})$ is the thermalization efficiency derived from the Rankine–Hugoniot jump condition \citep{2007ApJ...669..729K}.
We direct the reader to \citet{2015MNRAS.446.3992S} for more detailed information. 

\begin{figure*}
 \centering
 \includegraphics[width=1.9\columnwidth]{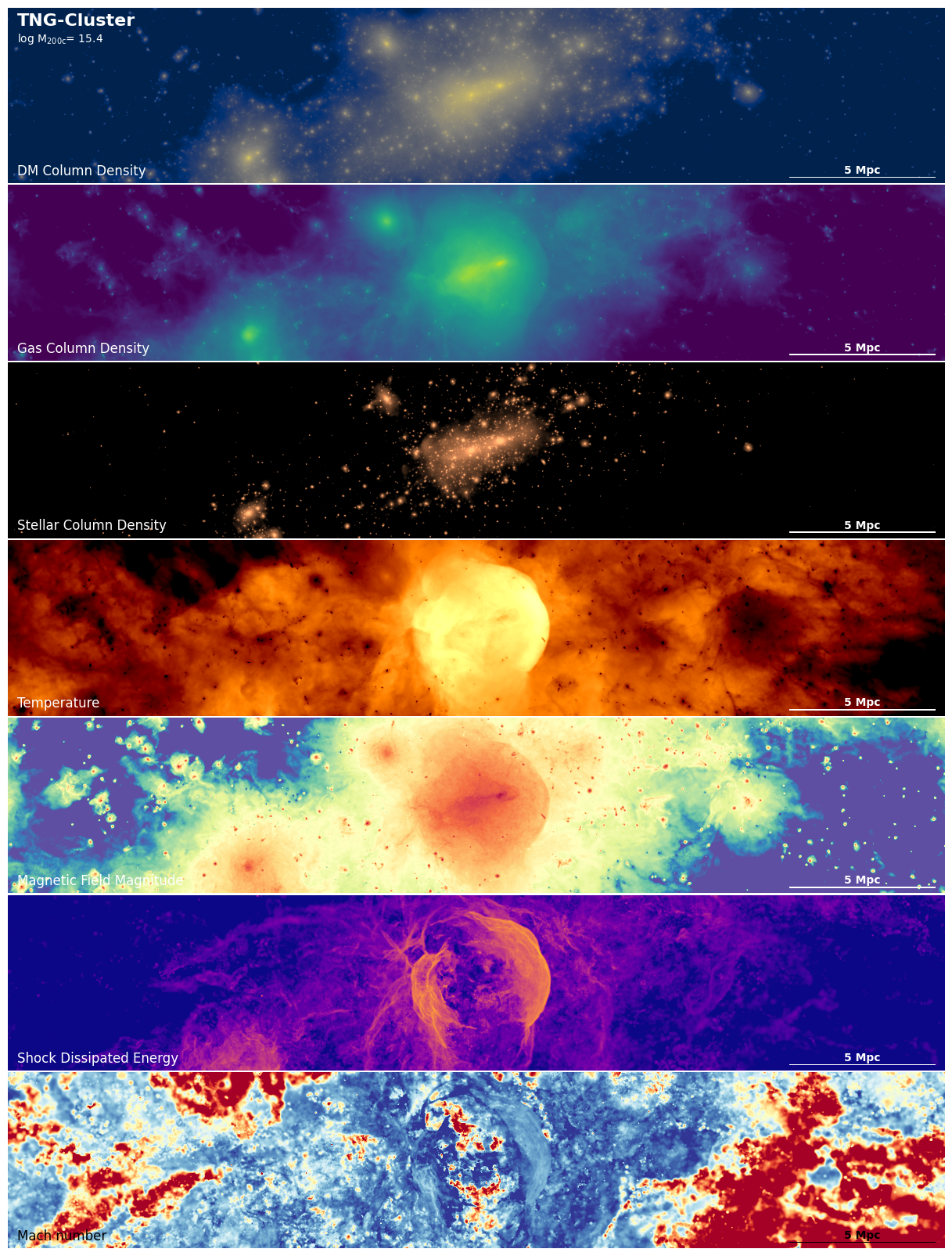}
 \caption{Most massive cluster merger in the TNG-Cluster simulation at $z=0$ ($\log M_{\rm 200c} = 15.4$). From top to bottom, each panel displays the projection of dark matter column density ($[10^{6.5},10^{9.5}]\rm~\MSUN~kpc^{-2}$), gas column density ($[10^{5.5},10^{8.5}]\rm~\MSUN~kpc^{-2}$), stellar column density ($[10^{4.5},10^{7.5}]\rm~\MSUN~kpc^{-2}$), density-weighted temperature ($[10^{6.5},10^{8.5}]\rm~K$), mass-weighted magnetic field strength ($[10^{-3},10]\rm~\mu G$), dissipated energy ($[10^{37},10^{42}]\rm~erg~s^{-1}~kpc^{-2}$), and the dissipated energy-weighted Mach number ($[2,50]$). 
 The map size is $32\rm~Mpc\times6\rm~Mpc$, the considered depth is 20 Mpc, and the colormaps of the images are log-scaled. 
 This figure demonstrates that TNG-Cluster can realize an extremely massive cluster merger and its 4~Mpc-size merger shocks while simultaneously modeling and resolving its environment in great detail. }
 \label{fig: Halo3}
\end{figure*}

We utilize these shock properties to estimate the radio emission of the cosmological shockwaves in the TNG-Cluster systems. The underlying assumption is that collisionless shockwaves can accelerate thermal plasma to the nonthermal regime via DSA, and the nonthermal plasma then produces synchrotron radio emission in the presence of the $\mu G$-scale intracluster magnetic fields.
The detailed acceleration process is still a subject of debate. In the following, we adopt the analytic model described by \citet{2007MNRAS.375...77H}, which in turn can be directly implemented using the simulation outputs in post-processing. Here, we summarize the salient aspects of our implementation.

Under the DSA process, the repetitive shock-crossing energizes the thermal particles to a relativistic regime.
The accelerated electrons at the shock fronts follow a power-law energy spectrum, 
\begin{equation}
    n_e(E)\propto E^{-s},
\end{equation}
where the energy spectral slope $s$ is defined as $s=(\sigma+2)/(\sigma-1)$ with $\sigma$ representing the shock compression rate. The shock compression rate is, in turn, a function of the shock strength and can be expressed as: $\sigma=(\gamma+1)\mathcal{M}^2/((\gamma-1)\mathcal{M}^2+2)$.
The energy spectrum truncates at an upper limit, denoted $E_{\rm max}$. This upper limit is the energy level at which the cooling offsets the energy gains from the single shock crossing \citep[][]{2003ApJ...585..128K,2011JKAS...44...49K}.
The truncated energy spectrum is 
\begin{equation}
    n_e(E)\propto E^{-s} \left(1-\frac{E}{E_{\rm max}}\right)^{s-2}.
\end{equation}

Following acceleration, CRe advect downstream and efficiently dissipate their energy through synchrotron radiation and inverse-Compton scattering \citep[][]{1962SvA.....6..317K}. As these energy loss rates are proportional to $E^{2}$, in practice, the energy spectrum is truncated at lower energies and is modified as 
\begin{equation}
    n_e(E,t)\propto E^{-s} \left[1-\left(E_{\rm max}^{-1} + \frac{C_{\rm cool}t}{m_{\rm e} c^2}\right)E\right]^{s-2}.
\end{equation}
Here, $t$ is the time since the electron acceleration and $C_{\rm cool}$ is the cooling coefficient defined as
\begin{equation}
    C_{\rm cool} = \frac{\sigma_{\rm T}}{6m_e c\pi} \left(B^2 + B^2_{\rm CMB}\right),
\end{equation}
where $\sigma_{\rm T}$ is the Thompson scattering cross section, $B$ is the strength of the intracluster magnetic fields, and $B_{\rm CMB}$ is the magnetic field strength that corresponds to the radiative energy density of the cosmic microwave radiation, which in turn follows $B_{\rm CMB}\sim3.24~\rm {\mu G} (1+z)^2$.

We assume that a constant fraction $\xi$ of the thermally dissipated energy transfers to nonthermal plasma energy.
Then, we can normalize the CRe energy spectrum with 
\begin{equation}
   \xi  E_{\rm diss} = Av_2 \int^{E_{\rm max}}_{E_{\rm min}}~dE~n_{\rm e} (E,0) E,
\end{equation}
where $E_{\rm min}$ is the minimum energy of the electrons involved in the acceleration process. Here, we use $E_{\rm max}=10^{10}m_{\rm e}c^2$ and $E_{\rm min}=0.01m_{\rm e}c^2$, following previous studies \citep[e.g.,][]{2007MNRAS.375...77H,2015ApJ...812...49H}.
The total synchrotron radiation emitted by the CRe population is given by integrating the energy spectrum over both energy and time,
\begin{equation}
    \frac{dP (\nu_{\rm obs},t)}{dA~d\nu} = \int^{\infty}_{0} dy \int^{E_{\rm max}}_{E_{\rm min}} dE n_{\rm e} (E,t) P_{\rm e} (\nu_{\rm obs},~\gamma_{\rm L}),
\end{equation}
where $\nu_{\rm obs}$ is the observed frequency, $y$ is the advection distance from the shock front that satisfies $y=v_2 t$, and $P_{\rm e}$ is the synchrotron power of a single electron:
\begin{equation}
    P_{\rm e} = \frac{\sqrt{3}\pi e^3 B}{4 m_e c^2} F(\nu_{\rm obs}/\nu_{\rm c}).
\end{equation}
In the above expression, $\nu_{\rm c}=(3\gamma_{\rm L}^2eB\sin{\theta})/(4\pi m_e c)$ is the characteristic frequency defined with the Lorentz factor $\gamma_{\rm L}$, and $F(x)\equiv\int^\infty_x d\xi K_{5/3}(\xi)$ (where $K_{5/3}(x)$ is a modified Bessel function). 
Combining these equations, we can derive the following analytic model using the gas and shock properties simulated by TNG-Cluster:
\begin{equation}
\label{eq:radio_power}
\begin{split}
    \frac{dP(\nu_{\rm obs})}{d\nu} = &~5.2\times10^{23}~\rm{W~Hz^{-1}} \times \left(\frac{\xi}{0.05}\right) \left(\frac{E_{\rm diss}}{10^{44} \rm erg~s^{-1}}\right)  \\
                     & ~~\frac{\left(\frac{B}{\rm \mu G}\right)^{\frac{s}{2}+1}}{\left(\frac{B}{\rm \mu G}\right)^2 +\left(\frac{B_{\rm CMB}}{\rm \mu G}\right)^2} \left(\frac{\nu_{\rm obs}}{1.4\rm~GHz}\right)^{-s/2} \Phi(\mathcal{M}),
\end{split}
\end{equation}
where $\Phi(\mathcal{M})$ describes the dependence on Mach number, with $\Phi(\mathcal{M}\rightarrow\infty)=1$ and steeply dropping for $\mathcal{M}<3$. 

As discussed in \citet{2007MNRAS.375...77H}, the normalization factor $5.2\times10^{23}~\rm{W~Hz^{-1}}$ represents an upper limit for the radio emission given the specified shock properties and magnetic field strength. 
For simplicity, we assume throughout a fixed efficiency of $\xi = 0.05$. Combined with the dissipation efficiency, this corresponds to an acceleration efficiency where $\mytilde0.8\%$ and $\mytilde5\%$ of the kinetic energy flux from $\mathcal{M}=2$ and $5$ merger shocks, respectively, are transferred to the CRe.
This assumption may overestimate the radio power of weak Mach number shocks \citep[e.g.,][]{2013MNRAS.435.1061P}.

We estimate the radio emissivity of the simulated merging clusters using 
Equation~\ref{eq:radio_power} and the identified shockwaves. Our method assumes that properties such as dissipation rate, shock strength, and magnetic field strength are constant within the CRe cooling timescale ($\mytilde0.1\rm~Gyr$).
We assign the radio emissivity to the shock surface cells.
By construction, the resulting radio emission will determine the minimum width of the radio relic, as the radio emission from the downstream is integrated on the shock surface. 
For example, at the observed frequency of $\nu_{\rm obs}=1.4\rm~$GHz, the radio emission from a shockwave with a Mach number of $\mathcal{M}\sim3.5$, advection velocity of $500\rm~km~s^{-1}$ and an intracluster magnetic field of $0.5\mu G$ at $z=0$ will drop by one order of magnitude at a distance of $\mytilde50\rm~kpc$  from the shock surface. We note that TNG-Cluster does resolve the downstream shock region, however we do not have the integrated history of the energy losses. We therefore use the dissipated energy and the magnetic field magnitude of the shock surface cells.

In summary, we compute the radio emissivity at the observed frequency with an analytic equation that operates directly on the TNG-Cluster simulation output in post-processing.
Radio emission is assigned to the shock surface cells. These cells will present a thin radio feature that does not account for the broader width of the feature due to cooling CRe moving downstream. 
We perform the calculation on the 8 full snapshots at $z\leq1$. 

We note that the shock finder also identifies the shockwaves generated by galactic feedback within and around galaxies. 
As these shocks inject large amounts of energy into a small volume, our modeling also produces point-like radio sources with high luminosity. We discuss this in \textsection~\ref{sec: extract}. We have also verified that the shock-dissipated energy and the radio morphologies reported in the current paper are well converged at the resolution of TNG-Cluster (see Appendix \ref{sec: app-res}). 

\subsection{X-ray emissivity in post-processing}

We derive the X-ray emissivity from the thermal plasma using the Astrophysical Plasma Emission Code (APEC) model \citep{2001ApJ...556L..91S}, which provides a table of line and continuum emissivities from the collisionally ionized, optically thin plasma. We utilized this table, along with the electron abundance and the metallicity obtained from the galaxy formation model, to estimate the net X-ray emissivity as a function of the energy band. 
In the following, we show results for the observed soft X-ray energy band by integrating the modeled X-ray emissivity within $0.5-2.0\rm~keV$.
We apply this only to gas cells within the high-resolution zoom-in region that meet the conditions $T>10^{6}\rm~K$ and $n_{\rm H}<10^{-1}\rm~cm^{-3}$ \citep{nelson2019a}. 

\begin{figure}
\centering
\includegraphics[width=\columnwidth]{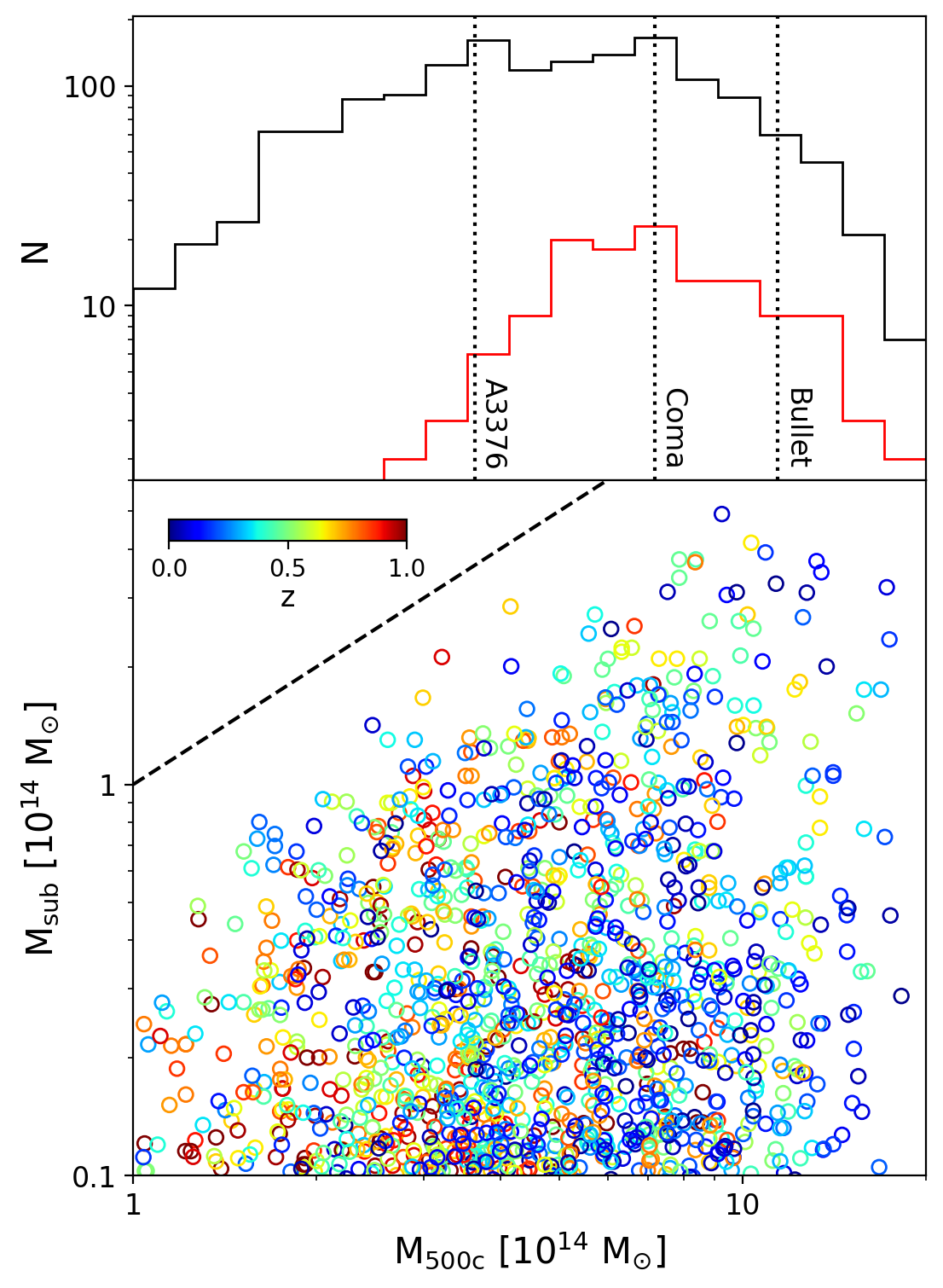}
 \caption{Properties of the cluster mergers realized and identified in the TNG-Cluster suite. We plot the distribution of the main cluster mass (top), and the main cluster mass vs. the subcluster mass (bottom). We include all merging galaxy clusters identified in the 50 snapshots at $z\leq1$.
 The $M_{\rm 500c}$ mass of the cluster is measured at the moment of the cluster collision. The mass of the merging companion, $M_{\rm sub}$, denotes its peak mass prior to the effects of stripping inside the main cluster potential (Section~\ref{sec:merger_def}). Thanks to the large sample and broad mass coverage, TNG-Cluster presents simulation counterparts to well-known cluster mergers, such as Abell 3376, Coma, and the Bullet Cluster.}
 \label{fig: mergers_stats}
\end{figure}

\begin{figure*}
 \centering
 \includegraphics[width=2\columnwidth]{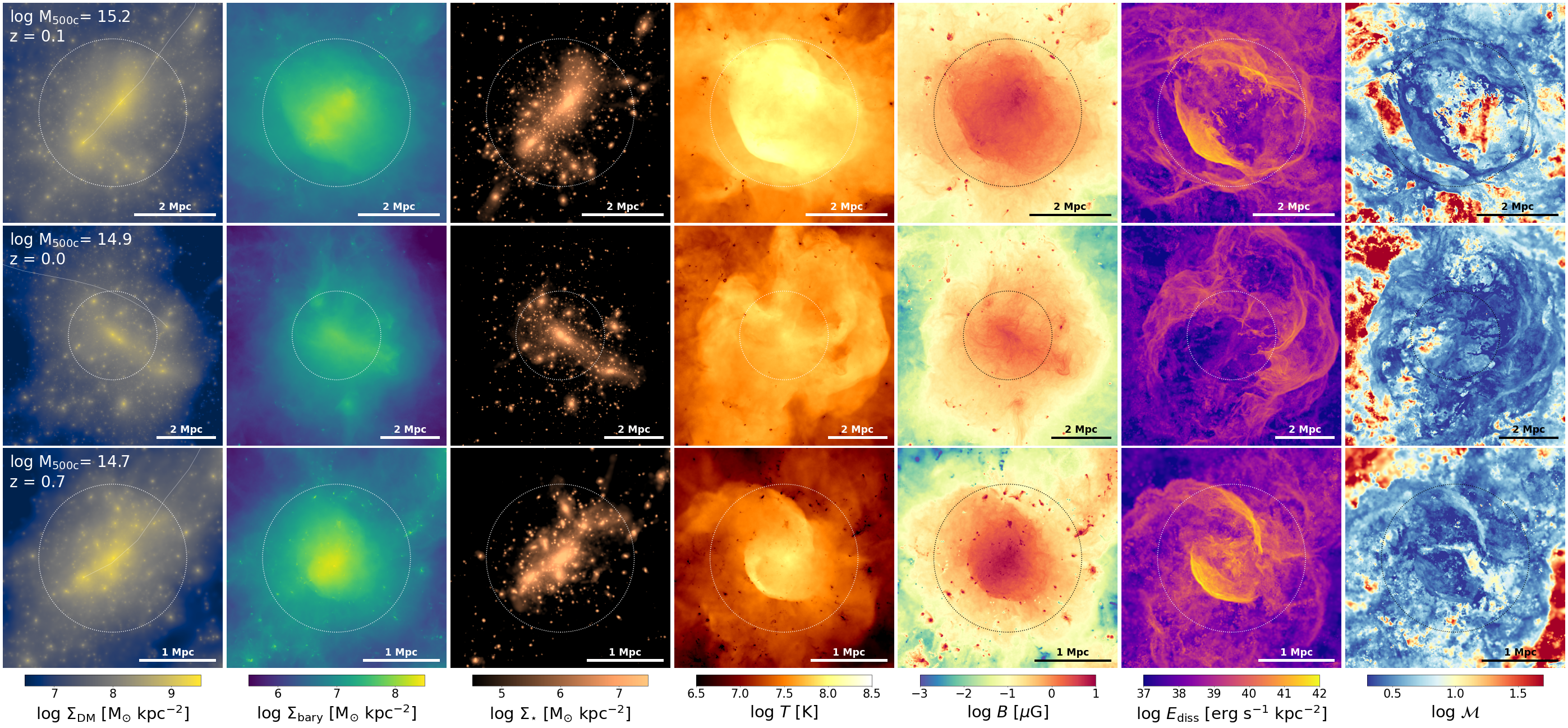}
 \caption{Projection maps of the same quantities as in Figure~\ref{fig: Halo3} of three mergers of the TNG-Cluster clusters with extreme merger properties.
 The considered depth is 20 Mpc.
 From top to bottom, we show a) a high-velocity collision with $\mytilde4300\rm~km~s^{-1}$, b) an off-axis cluster merger with pericenter separation $\mytilde1.8\rm~Mpc$, and c) a high redshift ($z=0.7$), massive ($\log~M_{\rm 200c}=14.8$) cluster merger. In all panels, the circles denote $R_{\rm 500c}$; the white curve in the DM density maps marks the path of the colliding subcluster over the last $\mytilde3\rm~Gyrs$. 
 }
 \label{fig: mergers}
\end{figure*}

\section{Cluster mergers \& radio relics}
\label{sec:result}

\subsection{The poster child of TNG-Cluster massive mergers}
\label{sec: merger_example}

Figure~\ref{fig: Halo3} presents the most massive merging cluster at $z=0$ in TNG-Cluster ($M_{\rm 200c}=2.7\times10^{15}\rm~\MSUN$, Halo $\rm ID~3$). This halo epitomizes the strengths of the TNG-Cluster simulation and sets the stage for our investigation of the merging population. 

In this system, two clusters with a total mass of $1.2\times10^{15}~\MSUN$ (left) and $3\times10^{14}~\MSUN$ (right of the image center), respectively, have experienced their first pericenter passage $\mytilde0.87\rm~Gyr$ ago with a non-zero pericenter separation ($\mytilde300\rm~kpc$) and a collision velocity of $\mytilde3400\rm~km~s^{-1}$ (in Figure~\ref{fig: Halo3}, the collision axis is nearly horizontal).
After the collision, the dark matter, gas, and stars of the clusters have dispersed along the collision axis. Meanwhile, the merger shocks, which are identifiable by the sharp temperature discontinuities, stretch in a direction perpendicular to this axis. 
The size of the shock front is gigantic, with the LLS of the merger shock reaching $\mytilde4.6\rm~Mpc$.

The TNG-Cluster setup allows us to realize and resolve the environment of this massive cluster merger in great detail.
At the cluster center, both the mixing processes around the cool core and the turbulent intracluster magnetic fields are resolved.
In addition, the simulation well resolves numerous cluster galaxies. 
The main cluster shown in Figure~\ref{fig: Halo3} hosts $\mytilde400$ galaxies with stellar mass $M_{\star}>10^{10}\rm~\MSUN$ within $2R_{\rm 200c}$.
A few of these satellites contain a cold magnetized gaseous disk, which highlights the scientific potential of TNG-Cluster in tracking the evolution of the cluster galaxies -- see \citet{2023arXiv231106337R} for more details on the evolution of satellite galaxies and their gas reservoirs.

The on-the-fly shock finder detects the complex nature of cosmological shockwaves around the merging cluster.
It reveals and distinguishes the two main classes of shockwaves identified in previous studies: weaker shocks ($\mathcal{M}\lesssim5$) and more energetic shockwaves near the cluster center, referred to as merger shocks, as well as stronger shocks ($\mathcal{M}\gtrsim10$) with lower energy density that encompasses large-scale structures, referred to as accretion shocks \citep[e.g.,][]{2003ApJ...593..599R,2008ApJ...689.1063S}. 
Moreover, although the merger shocks are $\mytilde2\rm~Mpc$ away from the cluster center, the shockwaves are resolved with a high spatial resolution. This resolution enables us to measure the magnetic field strength and shock strength fluctuations across the shockwaves. 

\subsection{Merging clusters from TNG-Cluster}
\label{sec: merger_lib}

In addition to the illustrative example shown in Figure~\ref{fig: Halo3}, the TNG-Cluster suite provides thousands of merging clusters with a wide range of masses, as we characterize in Figure \ref{fig: mergers_stats}.
By inspecting the 50 simulation snapshots available between $z=0$ and $z=1$, we have identified $\mytilde1800$ collisions between main clusters with a total mass $M_{\rm 500c}>10^{14}\rm~\MSUN$ and group-scale halos ($>10^{13}\rm~\MSUN$), as well as $\mytilde150$ collisions involving cluster-scale halos ($>10^{14}\rm~\MSUN$).
Furthermore, thanks to its halo selection strategy and large parent box size, TNG-Cluster also provides a large number of extremely massive merging clusters.
There are $\mytilde170$ mergers with the main cluster mass exceeding $M_{\rm 500c}>10^{15}\rm~\MSUN$, among which $\mytilde30$ are cluster-cluster mergers, with the subcluster mass being $M_{\rm sub}>10^{14}\rm~\MSUN$.
The wide mass distribution of the TNG-Cluster merger sample allows us to identify analogs to some well-known real mergers, such as Abell 3367 ($M_{\rm 500c}\sim3.6\times10^{14}~\MSUN$), the Coma cluster ($M_{\rm 500c}\sim7.2\times10^{14}~\MSUN$), and the Bullet cluster ($M_{\rm 500c}\sim11.4\times10^{14}~\MSUN$).

TNG-Cluster provides rare mergers with exceptional properties.
Figure~\ref{fig: mergers} showcases three such examples. 
The top row shows a high-velocity cluster merger where two clusters with masses of $1.3\times10^{15}~\MSUN$ and $2.8\times10^{14}~\MSUN$ collide with a velocity of $\mytilde4300\rm~km~s^{-1}$. 
As a result of the high-velocity collision, the ICM is heated to $\gtrsim10^{8}\rm~K$, and two energetic yet weak ($\mathcal{M}\sim3$) merger shocks are propagating along the collision axis. At the displayed merger phase, the colliding subcluster is at a distance of
$\mytilde1.4\rm~Mpc$ from the main cluster, and it has been $\mytilde0.4\rm~Gyr$ since the collision. 
It should be noted that both the collision velocity and the combined mass ($M_{\rm 500c}=1.7\times10^{15}~\MSUN$) of this system are comparable to the shock velocity and the mass ($M_{\rm 500c}=1.1\times10^{15}~\MSUN$) of the Bullet Cluster \citep{2004ApJ...606..819M}. 
As the extreme collision velocity of the Bullet has been considered a challenge to the $\Lambda$CDM paradigm \citep[e.g.,][]{2010ApJ...718...60L,2015JCAP...04..050K}, the presence of these high-velocity cluster mergers in TNG-Cluster can serve as a valuable reference for interpreting the system and forecasting the time evolution of mergers with Bullet-like characteristics.

The middle row of Figure~\ref{fig: mergers} shows a merger between a $6~\times10^{14}~\MSUN$ main cluster and a $\mytilde6~\times10^{13}~\MSUN$ subcluster with a large pericenter separation of $\mytilde1.8\rm~Mpc$. 
This off-axis cluster-group merger generates a merger shock ahead of the merging companion with relatively low dissipated energy compared to the bullet-like near head-on collision. 
Considering both the collision velocity of $\mytilde2,000\rm~km~s^{-1}$ and the substantial pericenter separation, this minor merger will inject a large amount of angular momentum into the cluster environment and generate a rotational ICM motion that can persist for billions of years \citep[e.g.,][]{2006ApJ...650..102A,2010ApJ...717..908Z,2011MNRAS.413.2057R}. 
TNG-Cluster contains $\mytilde100$ cluster mergers with a pericenter separation $\gtrsim1\rm~Mpc$, enabling statistical analysis of rotational motion driven by cluster mergers.

Finally, the TNG-Cluster simulation provides high-redshift massive cluster mergers. We show an example in the bottom row of Figure~\ref{fig: mergers}, where a cluster merger with $M_{\rm 200c}\sim6\times10^{14}~\MSUN$ is found at $z=0.7$. This binary cluster merger with a relatively low mass ratio $M_{\rm main} /M_{\rm sub} \sim4$ generates two merger shocks with dissipated energy comparable to that in the low-redshift massive cluster mergers. 
Massive cluster mergers at high redshift have also been claimed as a possible challenge to the $\Lambda$CDM cosmology \citep[e.g.,][]{2014ApJ...785...20J,2019ApJ...887...76K,2021ApJ...923..101K,2021MNRAS.500.5249A}, and these high-redshift TNG-Cluster mergers provide simulation counterparts that can be used to interpret such observations. We note that this example (the bottom panel Figure~\ref{fig: mergers}) is not even the most massive merging cluster. For instance, the most massive cluster mergers at $z=0.7$ and $z=1.0$ have total masses of $M_{\rm 200c}=1.1\times10^{15}~\MSUN$ and $9.7\times10^{14}~\MSUN$, respectively.

\begin{figure*}
 \centering
 \includegraphics[width=2\columnwidth]{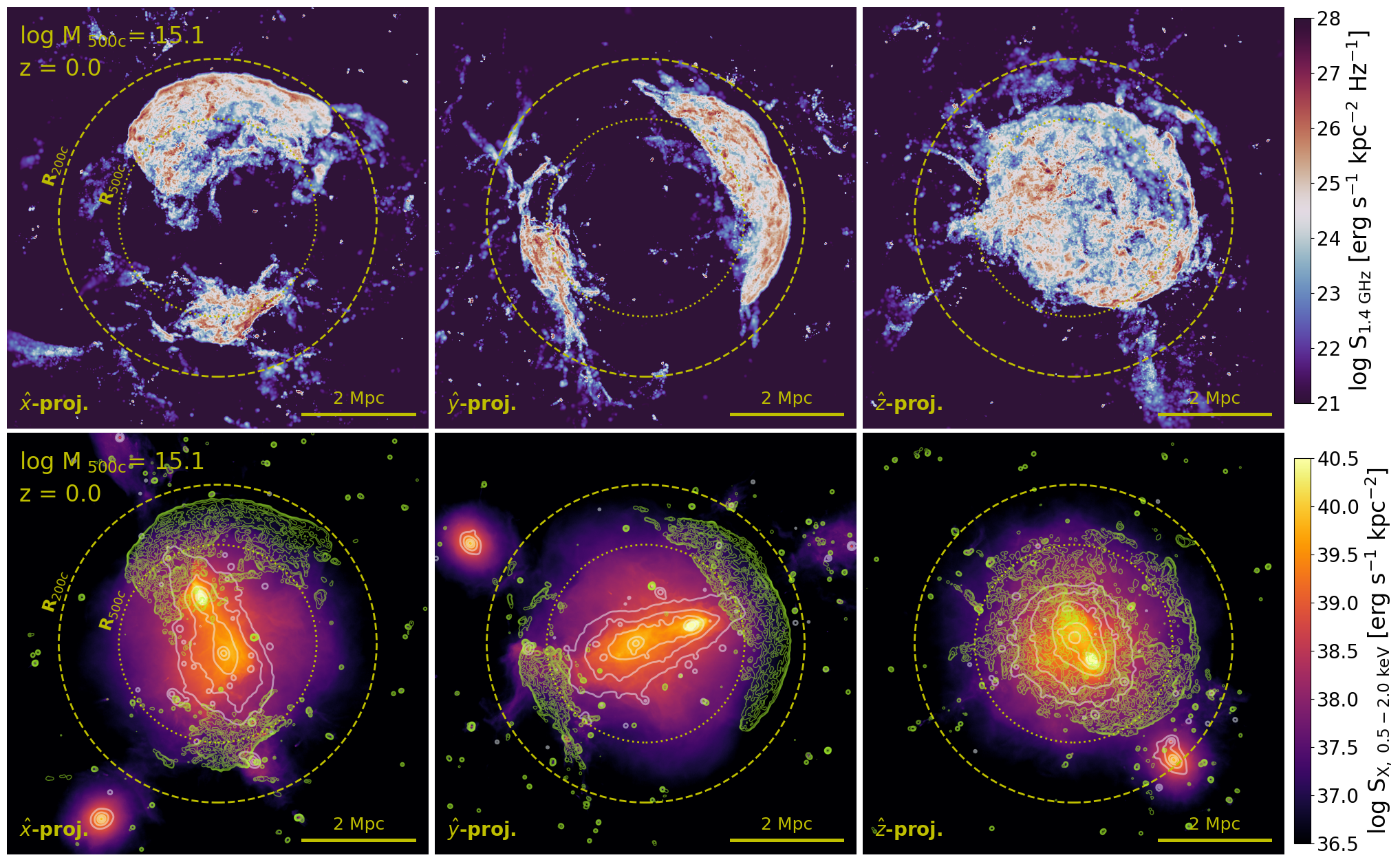}
 \caption{Radio surface brightness map (top) and multiwavelength map (bottom) in three different projections of the same TNG-Cluster merging cluster as in Figure~\ref{fig: Halo3}. The yellow circles mark $R_{\rm 500c}~(1.7\rm~Mpc)$ and $R_{\rm 200c}~(2.8\rm~Mpc)$ from the cluster center.
 The green and white contours on the X-ray map depict radio surface brightness and mass surface density, smoothed with $\sigma=10$ and $20\rm~kpc$ Gaussian kernels and spaced equally in logscale in between $[10^{24},10^{27}]\rm~erg~s^{-1}~Hz^{-1}~kpc^{-2}$ and $[10^{8.3},10^{9.3}]\rm~\MSUN~kpc^{-2}$, respectively. 
 The clearly visible double radio relics are elongated tangentially with respect to the mass center and their LLS measures $\mytilde4\rm~Mpc$. Moreover, they form a nonuniform radio halo-like feature in the z-axis projection that fills the cluster volume to $R_{500\rm c}$. The total radio power of these structures is $P_{1.4\rm~GHz}\sim10^{25}~\rm W~Hz^{-1}$.}
 \label{fig: Halo3_radio}
\end{figure*}

\subsection{An archive of simulated radio relics}

\label{sec: relic_lib}

\begin{figure*}
 \centering
 \includegraphics[width=2\columnwidth]{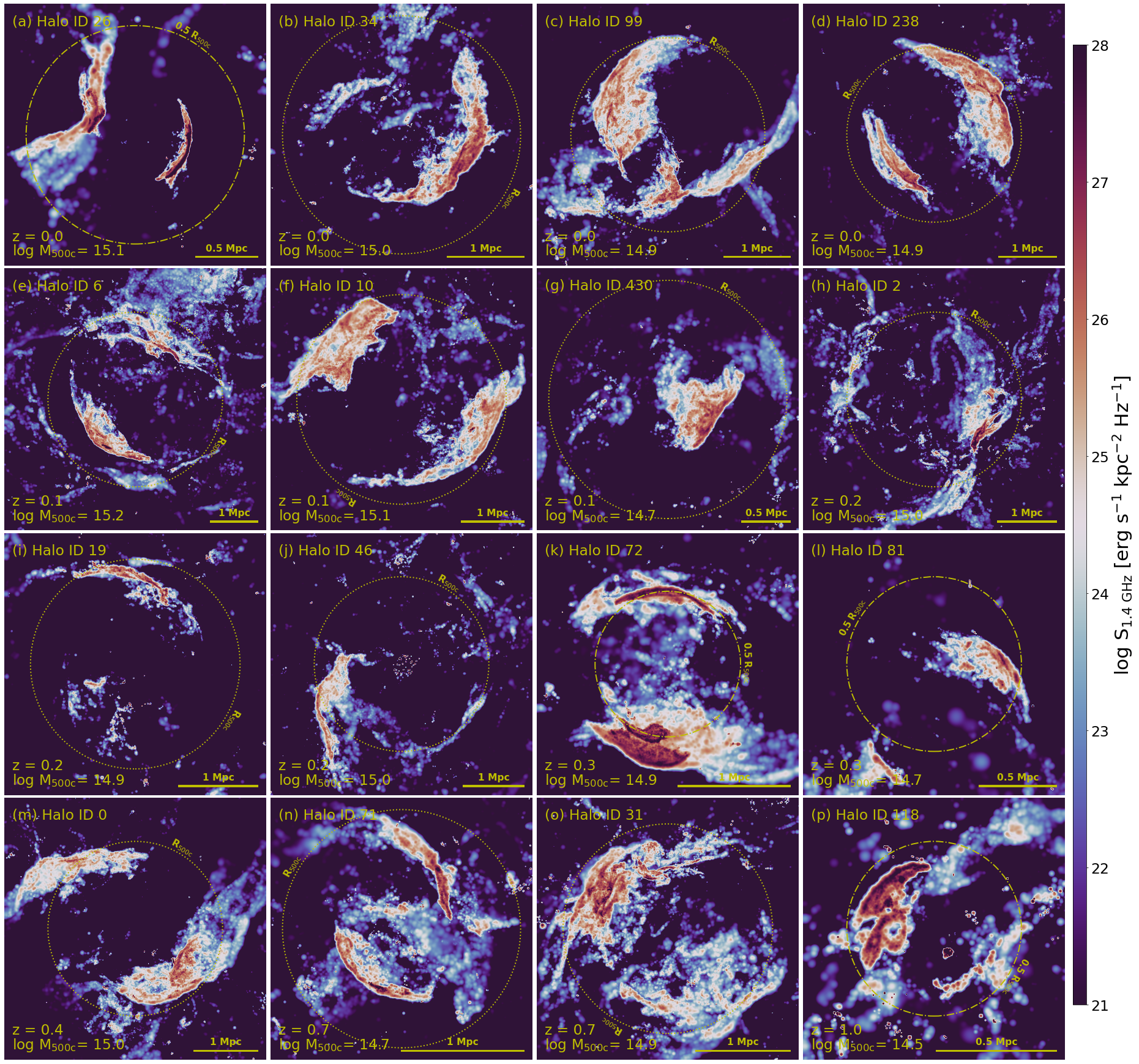}
 \caption{Radio surface brightness maps of 16 radio relic systems in the TNG-Cluster simulation that showcase a diversity of features and configurations. 
 The color scale of the radio surface brightness maps ranges from $10^{21}$ to $10^{28}\rm~erg~s^{-1}~Hz^{-1}~kpc^{-2}$. 
 The dotted and dash-dotted yellow circles mark $R_{\rm 500c}$ and $0.5R_{\rm 500c}$, respectively. 
 TNG-Cluster radio relics present a diverse morphology, including double, single, and linear or inverted radio relics.
 }
 \label{fig: reliczoo}
\end{figure*}

By applying the post-processing radio calculations described in Section~\ref{sec: emissivity} to the TNG-Cluster output, we find that thousands of merging clusters exhibit a wide variety of radio relics that resemble those observed. In the following, we provide a qualitative description of the radio relics found in TNG-Cluster, and we have made their images and data publicly available with this paper for future use by the community. 

\subsubsection{Examples of double radio relics from TNG-Cluster}

Figure~\ref{fig: Halo3_radio} presents the radio maps in three different projections of the same massive cluster merger shown in Figure \ref{fig: Halo3}. 
The cluster shows double radio relics in the $x$- and $y$-axes projections, which extend tangentially to the cluster center. 
It should be noted that the radio relics are extremely large: the larger radio relic is propagating ahead of the subcluster and has an LLS of $\mytilde4.1\rm~Mpc$ with a thickness of $\mytilde0.6\rm~Mpc$; the smaller radio relic, adjacent to the main cluster, is $\mytilde2.5\rm~Mpc$ long and $\mytilde0.6\rm~Mpc$ wide.
Thanks to the relatively uniform surface brightness, we expect that the full extent of these radio relics can be observed in radio observations with $\mytilde0.05\rm~\mu Jy~''^{-2}$ sensitivity. 

\begin{figure*}
 \centering
 \includegraphics[width=2\columnwidth]{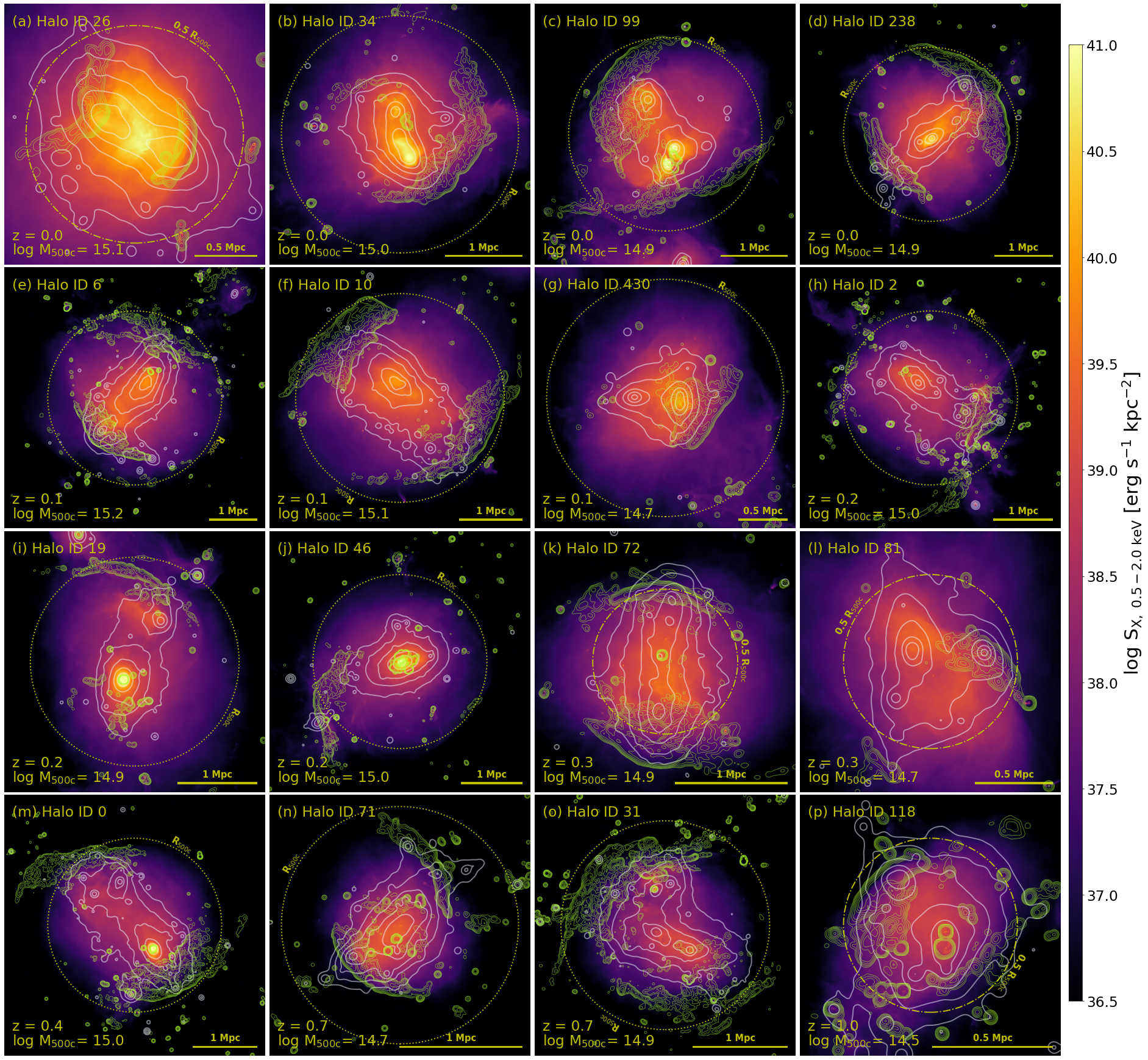}
 \caption{Multiwavelength projection maps as in Figure~\ref{fig: Halo3_radio} of the merging clusters presented in Figure~\ref{fig: reliczoo}. The majority of the systems present elongated mass and X-ray morphology, and the radio relics extend perpendicular to these elongations.}
 \label{fig: mw_relic_zoo}
\end{figure*}

As is evident in Figure~\ref{fig: Halo3_radio}, the morphology of the radio features varies with the projection axes. 
The double radio relics present an arc-like morphology in the $x$- and $y$-axes projections. However, the $z$-axis projection shows a halo-like morphology. Note that in its physical origin and 3D extent, it is different from the classical radio halos formed by turbulent reacceleration \citep[e.g.,][]{2001MNRAS.320..365B} or secondary electrons from hadronic interactions \citep[e.g.,][]{2008MNRAS.385.1211P}, which are volume-filling.
This serves as an important reminder that caution is needed when classifying radio features when the merger is viewed along the collision axis \citep[see also ][]{2013ApJ...765...21S}. We discuss this issue in Section~\ref{sec: halo}.
Comparing the radio relic morphologies with the dark matter distributions between the $x$- and $y$-projections reveals that the double radio relics in the $y$-projection are more symmetric with respect to the merger axis.

On larger scales around this system, there are no bright radio features that can be attributed to cosmological accretion shocks.
In Figure~\ref{fig: Halo3}, it is evident that this merging cluster is surrounded by multiple layers of shockwaves connected to nearby large-scale structures. We can identify some localized patches of radio emission around the radio relics, which might spatially be associated with these shockwaves. However, the power of these radio features is already three orders of magnitude lower than that of the radio relics, making the accretion features barely noticeable, even if they are indeed from accretion shocks.
This suggests that accretion shocks may have a negligible impact on the morphology of radio relics.  

The bottom row of Figure~\ref{fig: Halo3_radio} shows the X-ray luminosity maps overlaid on the surface mass density contours. These overlays show the spatial connection between the radio sources and the underlying matter distributions.
The total mass and the X-ray gas extend mostly along the $z$-axis (i.e., the collision axis), and the mass and X-ray surface brightness peaks coincide. 
Nonetheless, the relative alignment between the double radio relics and the mass or X-ray features varies with viewing angles. 
The centers of the radio relics are aligned with the two mass or X-ray peaks in the $y$-axis projection map, whereas the relic centers are misaligned with other features in the $x$-axis projection. 
We interpret the misalignment between the centers as the result of an off-axis collision.
The clusters have collided with a $\mytilde300\rm~kpc$ pericenter separation along the y-axis, which provides angular momentum and triggers a rotation of the dark halos and the ICM. In contrast, the merger shocks are launched along the collision axis and continue with relatively straight trajectories \citep[see also][]{2011MNRAS.418..230V,2019ApJ...874..112S,2020ApJ...894...60L}.

The multiwavelength maps also provide a chance to investigate the impact of an infalling massive satellite (not yet involved in the collision) on the radio relic morphology.
In the $x$- and $y$-axis projections, the radio relic in front of the main cluster coincides with the diffuse X-ray emission, where we can identify a $\mytilde3\times10^{13}~\MSUN$ halo, which is falling toward the cluster with a velocity of $\mytilde1,000\rm~km~s^{-1}$.
As the motion of the falling gas enhances the kinetic flux, the adjacent
radio relic can become brighter and may even warp \citep[e.g.,][]{2023ApJ...957L..16B}.
We discuss this process in more detail in Section~\ref{sec: Inverted}.

\subsubsection{Morphological diversity of TNG-Cluster radio relics}

The radio image in Figure~\ref{fig: Halo3_radio} provides a textbook-example symmetric double radio relic system. In Figure~\ref{fig: reliczoo}, we show 16 radio relics from the TNG-Cluster with diverse morphologies, including double and single radio relics. 
The projection axis is chosen to optimize the visualization of the arc-like morphology.\footnote{The radio maps of all the 352 re-simulated clusters are available online at \url{https://www.tng-project.org/cluster/gallery/}.}
Hereafter, we refer to halo ID as the halo IDs from the original parent volume \citep[see Table B1 of][]{2023arXiv231106338N}. 

Typically, the double radio relics extend tangentially with respect to the cluster center (e.g., panel~{\tt d}), whereas their detailed characteristics vary among the clusters.
The two radio relics of a double-relic system can at times have a comparable size and luminosity (e.g., panel~{\tt f}). However, more systems present asymmetry in size (e.g., panel~{\tt d}) and luminosity (e.g., panel~{\tt m}). 
Their thickness varies from thin (e.g., $<0.1\rm~Mpc$ in Halo 26, panel~{\tt a}) to thick ($\mytilde0.6\rm~Mpc$ in Halo 10, panel~{\tt f}).
Radio relics from the TNG-Cluster present a diverse range of curvatures as well. 
For example, the edge of the upper radio relic of Halo 10 in panel~{\tt f} exhibits a linear extension. In contrast, the left-side radio relic of Halo 26 in panel~{\tt b} shows a long tail and presents an inverted (convex toward the cluster center) radio relic.
We can also notice surface brightness fluctuations across the relic, and in the extreme case of panel~{\tt k}, we can even identify two layers of shock fronts inside a single radio relic. 
The majority of double radio relics bracket the cluster center, while an exception exists (panel~{\tt l}).
Both the high-velocity (panel~{\tt e}) and the high-redshift mergers (panel~{\tt n}) presented in Figure \ref{fig: mergers} exhibit double radio relics.

Single radio relic systems also present diverse and complex morphologies.
Halo 34 in panel {\tt b} shows a textbook arc-like single radio relic with a large extension ($\mytilde2\rm~Mpc$) and a thin width ($\mytilde0.2\rm~Mpc$).
On the other hand, Halo 430 in panel~{\tt g} possesses a short radio relic ($\mytilde0.5\rm~Mpc$), whose width is comparable to its extension.
We also find a single radio relic system, where the relic stretches toward the cluster periphery (panel~{\tt j}).
Similarly to double radio relics, among single radio relic systems, there is a case with an inverted radio relic (e.g., Halo 2, panel~{\tt h}) and a double-layered radio relic (e.g., Halo 118, panel~{\tt p}).

We highlight that, although the simulated radio relics closely follow the intrinsic morphology of the shockwaves by construction, they still present a diverse morphology, with asymmetric luminosity or size and even single radio relics.
This implies that such diverse relic morphology does not necessarily require inhomogeneous acceleration efficiency, but can originate solely from the complex nature of cosmological shockwaves. 
In real clusters, the acceleration efficiency can be inhomogeneous due to fossil CRe or different shockwaves, possibly further complicating the morphology of radio relics beyond that of the simulated radio relics presented here with a simple acceleration model. 

Figure~\ref{fig: mw_relic_zoo} displays the multiwavelength maps of the 16 radio relic systems presented in Figure~\ref{fig: reliczoo}.
In general, merging systems present extended mass distributions with multiple mass peaks. At the same time, the radio relics extend in the perpendicular direction with respect to the mass elongation in both double and single radio relic systems. 
The elongation of the diffuse X-ray emission roughly follows the mass distribution and thus extends in the direction perpendicular to the radio relic orientation. 
Nonetheless, in certain instances, the X-ray emission can be dissociated from the dark halo (e.g., panel~{\tt a}) or may lack a signal in the main or subcluster center (e.g., panel~{\tt m}). 

When more than two clusters are involved in mergers, the relation between the relic orientation and the apparent mass elongation becomes complicated.
For example, Halo 81 in panel~{\tt l} presents two dominant mass peaks along the $x$-axis, while the vector connecting its double radio relics is significantly tilted with respect to the $x$-axis.
By tracking the time evolution of this merger, we find that this system was, in fact, a triple merger. That is, the radio relics are the result of the collision between the second (right) and the third (lower left) most massive clusters (prior to the collision, the second most massive cluster current on the right side was located near the lower left corner). Thus, the shock normal of the radio relics is aligned with the collision axis, and thus still serves as a robust indicator of the collision axis even during this complex triple merger.

\subsubsection{TNG-Cluster analogs to observed systems}

Among the numerous radio relics produced by TNG-Cluster, we can visually identify several analogs to observed cases. 
For example, Halo 238 (panel~{\tt d} in Figures~\ref{fig: reliczoo} and \ref{fig: mw_relic_zoo}) exhibits double radio relics with an LLS of $\mytilde 2.6 \rm~Mpc$,  a discontinuous shock front, and a cool core in the extended X-ray map. These features resemble the observed signatures of Abell 3667 \citep[e.g.,][]{2022A&A...659A.146D}. 
Halo 2 (panel {\tt h}) is an analog to Coma \citep[e.g.,][]{1998A&A...332..395E}, where we can identify a $\mytilde0.8\rm~Mpc$-size inverted radio relic in the vicinity of the unrelaxed galaxy cluster. 
Halo 19 (panel~{\tt i}) showcases a radio relic that extends in the same direction as the X-ray tail, as observed in Abell 115 \citep[e.g.,][]{2016MNRAS.460L..84B,2019ApJ...874..143K}. Halo 46 (panel~{\tt j}) is similar to Abell 521 as it hosts an Mpc-size radio relic, which elongates toward the cluster periphery \citep[e.g.,][]{2008A&A...486..347G,2020ApJ...903..151Y}.  

We expect that these TNG-Cluster analogs can provide insight into their merger scenario responsible for forming the observed radio relics. 
For example, the Abell 115-analog, Halo 19 (panel~{\tt i}), is an off-axis cluster merger. 
In the image plane of Figure~\ref{fig: mw_relic_zoo}, the subcluster has fallen into the main cluster along the $y$-axis with a large pericenter separation of $\mytilde450\rm~kpc~(\mytilde0.3~R_{\rm 500c})$ toward the negative $x$-axis direction. 
The pressure stripping forms the gas tail, which rotates after the collision and thus its orientation becomes perpendicular to the merger axis.
In contrast, the radio relic maintains its orientation and stays perpendicular to the collision axis (the $x$-axis). Thus, the two features align at a particular epoch ($z=0.1$ in this case), similar to the case in Abell 115.
As demonstrated in this example, simulated counterparts of observed radio relic systems can provide plausible merger scenarios to explain the observed morphology, position, and orientation of radio relics.  
This can be further fine-tuned under a constrained cosmological setup or using idealized simulations that can reproduce the remaining details of the observations. 

\section{Statistical properties of TNG-Cluster radio relics}
\label{sec: statistics}

As showcased in the previous Sections, the TNG-Cluster simulation suite provides a large and rich archive of merging galaxy clusters and their radio relics. The latter exhibit diverse radio morphologies, including multiple analogs to observed systems. This indicates that the shockwaves of cosmological and hierarchical-growth origin simulated within TNG-Cluster can solely reproduce the observed radio relic morphological complexity.
In this Section, we perform a more quantitative and statistical analysis of TNG-Cluster radio relics to investigate how well the simulation predicts the observed properties and relations. 
We remind the readers that, unlike the morphology, the radio-relic luminosity is model-dependent and can be shifted by varying acceleration efficiency (see Section~\ref{sec: emissivity}). 

\begin{figure}
 \centering
 \includegraphics[width=0.92\columnwidth]{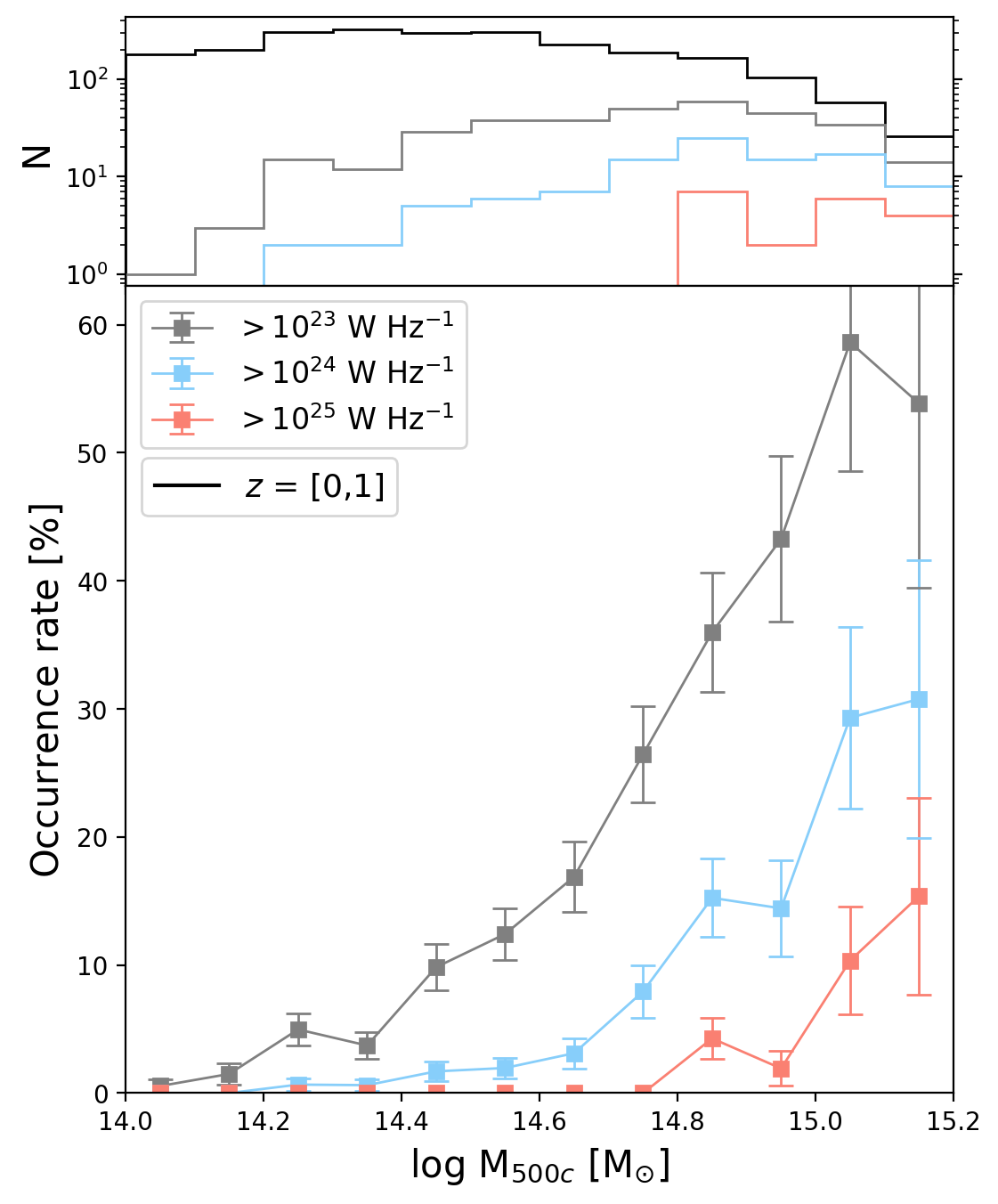}
 \includegraphics[width=0.9\columnwidth]{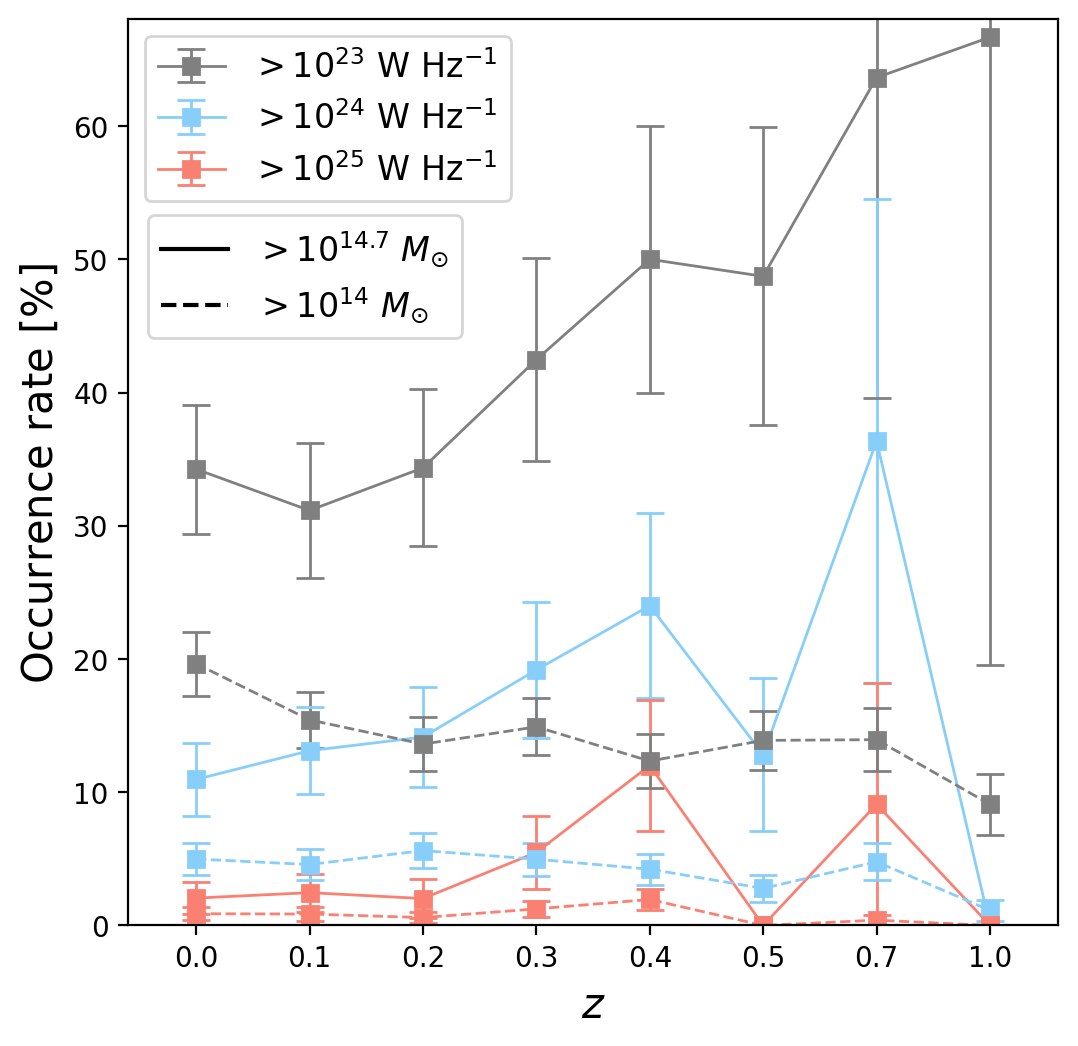}
 \caption{Occurrence rate of radio relics according to TNG-Cluster as a function of primary cluster mass (top) and redshift (bottom). The data points are colored based on different luminosity cuts, and the error bar shows the Poisson noise.
 The top subplot quantifies the number of clusters in each mass bin (black histogram) and those hosting radio relics above a certain radio luminosity (colored histograms). The final bin includes the three clusters with $M_{\rm 500c}>10^{15.2}~\MSUN$. 
 In the redshift distribution, the dotted and solid lines mark the clusters with $M_{\rm 500c}>10^{14}~\MSUN$ and $M_{\rm 500c}>5\times10^{14}~\MSUN$, respectively.
 The fraction of clusters that host radio relics increases with the system's mass, whereas the redshift dependence is less unclear.}
 \label{fig: fraction_mass}
\end{figure}

\subsection{Separation of relic cells from those with other origins}

\label{sec: extract}
The shockwaves produced during cluster mergers generate radio emissions and feature radio relics. 
However, in the TNG model, the AGN activity of the SMBHs in the cluster member galaxies can also generate shocks around the central galaxy and massive satellites.
The energy injected via AGN feedback creates shockwaves that expand outward more or less spherically, with possibly large Mach numbers and shock-dissipated energy comparable to that of merger shocks \citep[][]{2017MNRAS.465.3291W}. For example, SMBH-driven shocks produced in Milky Way and Andromeda-like galaxies in the TNG model are responsible for eROSITA and Fermi-like bubbles in their circumgalactic medium \citep{2021MNRAS.508.4667P}. In galaxy clusters and for the purposes of this paper, such AGN-driven shocks may result in bright radio point sources at the centers (see panel~{\tt j} in Figure~\ref{fig: reliczoo}) as well as extended spherical radio emissions. Therefore, a priori radio emissions within the TNG-Cluster are a mixture of merger shock-driven and AGN feedback-driven signatures, possibly in addition to shocks due to cosmological accretion or collapse, which do not often manifest in our maps (see previous Sections). 

To understand and compare with the observed properties of merger shock-driven radio emissions, we first need to identify and extract the cells contributing to the emission of radio relics.
First, we use the shock surface cells within $2R_{\rm 200c}$ and create a 3-dimensional, radio emissivity-weighted histogram with a bin size of $20\rm~kpc$. 
The shockwaves from cosmological gas accretion and collapse can overlap with low-energy cluster-merger shocks (see the dissipated energy map in Figure~\ref{fig: Halo3} and \ref{fig: mergers}). 
To avoid including all structures, we mask bins with radio luminosity $<10^{17}\rm~W~Hz^{-1}$ and group the unmasked bins based on their spatial connectivity.
Finally, we select the two brightest structures within the five groups with the largest volume as the radio relic candidates, assuming that merger shock fronts span a larger volume than the shockwaves from feedback. 

We derive the properties of radio relics using the cells included in each radio relic group (hereafter relic cells). 
We estimate the radio relic luminosity with the integrated luminosity of the relic cells and define the relic center as the radio luminosity-weighted center of the relic cells.
As weak features can cause an overestimation, we define LLS as the largest separation between the pair of relic cells whose radio luminosity exceeds $>10^{19}\rm~W~Hz^{-1}$.
For example, the sizes of the two radio relics in Halo 3 (Figure \ref{fig: Halo3_radio}) are estimated to be $\mytilde4.4\rm~Mpc$ and $\mytilde2.5\rm~Mpc$ based on this criterion. 
We note that the LLS measured in 3-dimensional space is an upper limit for the LLS in projection.

The procedure above removes AGN-like point sources. Nevertheless, extended radio-bright AGN shockwaves can overlap with faint cosmological shockwaves and merger shocks, and hence may be misinterpreted as bright radio relics. 
To minimize this contamination, we discard the structures with radio-luminosity weighted Mach number $\mathcal{M}_{\rm radio}>10$ or with the LLS  $\lesssim0.3\rm~Mpc$. 
In the end, we visually confirm the validity of our candidate radio relics.

\subsection{Abundance of TNG-Cluster radio relics} 
\label{sec: relic_count}

Based on the radio-relic identification described above and on the cluster mergers realized by TNG-Cluster, we find that the latter returns $\mytilde300$, $\mytilde100$, and $\mytilde20$ radio relics with radio luminosities $P_{\rm 1.4~GHz} >10^{23}\rm~W~Hz^{-1}$, $>10^{24}\rm~W~Hz^{-1}$, and $>10^{25}\rm~W~Hz^{-1}$, respectively, across 8 snapshots in the redshift range $z=0-1$.

The top main panel of Figure~\ref{fig: fraction_mass} presents the occurrence rate of $z=0-1$ radio relics depending on the mass of the primary merging cluster. More massive clusters have progressively larger chances of hosting radio relics. 
For example, $\mytilde14\pm1\%$ of clusters with $M_{\rm 500c} > 10^{14}~\MSUN$ host radio relics with $P_{\rm 1.4~GHz}>10^{23}\rm~W~Hz^{-1}$.
This rate of radio relic occurrence increases to $\mytilde38\pm3\%$ in massive clusters with $M_{\rm 500c} > 5\times10^{14}~\MSUN$.
The occurrence rate decreases with a higher luminosity cut, for example, $\mytilde4\pm0.4\%$ and $\mytilde15\pm2\%$ of $M_{\rm 500c} > 10^{14}~\MSUN$ and $ > 5\times10^{14}~\MSUN$ clusters present radio relics with $P_{\rm 1.4~GHz}>10^{24}\rm~W~Hz^{-1}$, respectively.
All radio relics with $P_{\rm 1.4~GHz}>10^{25}\rm~W~Hz^{-1}$ are found only in the massive clusters whose masses exceed $M_{\rm 500c} > 5\times10^{14}~\MSUN$, but are rare (occurrence rate of $\mytilde3\pm1\%$). 
Their merger shock origin can explain the higher occurrence rate of radio relics in massive clusters, as the relic luminosity increases with the cluster mass as more energy is released from the high-mass merger. 

We recall that TNG-Cluster focuses on massive clusters; therefore, the occurrence rates quantified here, for example, per mass bin, come from varying numbers of clusters from varying redshifts. 
For example, the clusters in the mass bin $\log M_{\rm 500c}=[15.0,15.2]$ are 81 in total and have an average redshift of $z\sim0.15~$ whereas the clusters in $\log M_{\rm 500c}=[14.0,14.2]$ have an average redshift of $z\sim0.52~(N=382)$.
To examine a possible redshift evolution effect, we compare the occurrence rate of radio relics as a function of the cosmic epoch in the bottom panel of Figure \ref{fig: fraction_mass}.  
When all clusters with $>10^{14}\MSUN$ are considered, the occurrence rate of radio relics with $P_{\rm 1.4~GHz}>10^{23}\rm~W~Hz^{-1}$ is consistently low at $\mytilde15\%$ across the redshift range, and only marginally decreases with redshift. 
For massive systems with $M_{\rm 500c} > 5\times10^{14}~\MSUN$, the trend is opposite, and the occurrence rate of bright radio relics appears to increase with redshift.
We further discuss and interpret the trends of Figure~\ref{fig: fraction_mass} in Section~\ref{sec: relic_stat}.

\begin{figure}
 \centering
 \includegraphics[width=0.75\columnwidth]{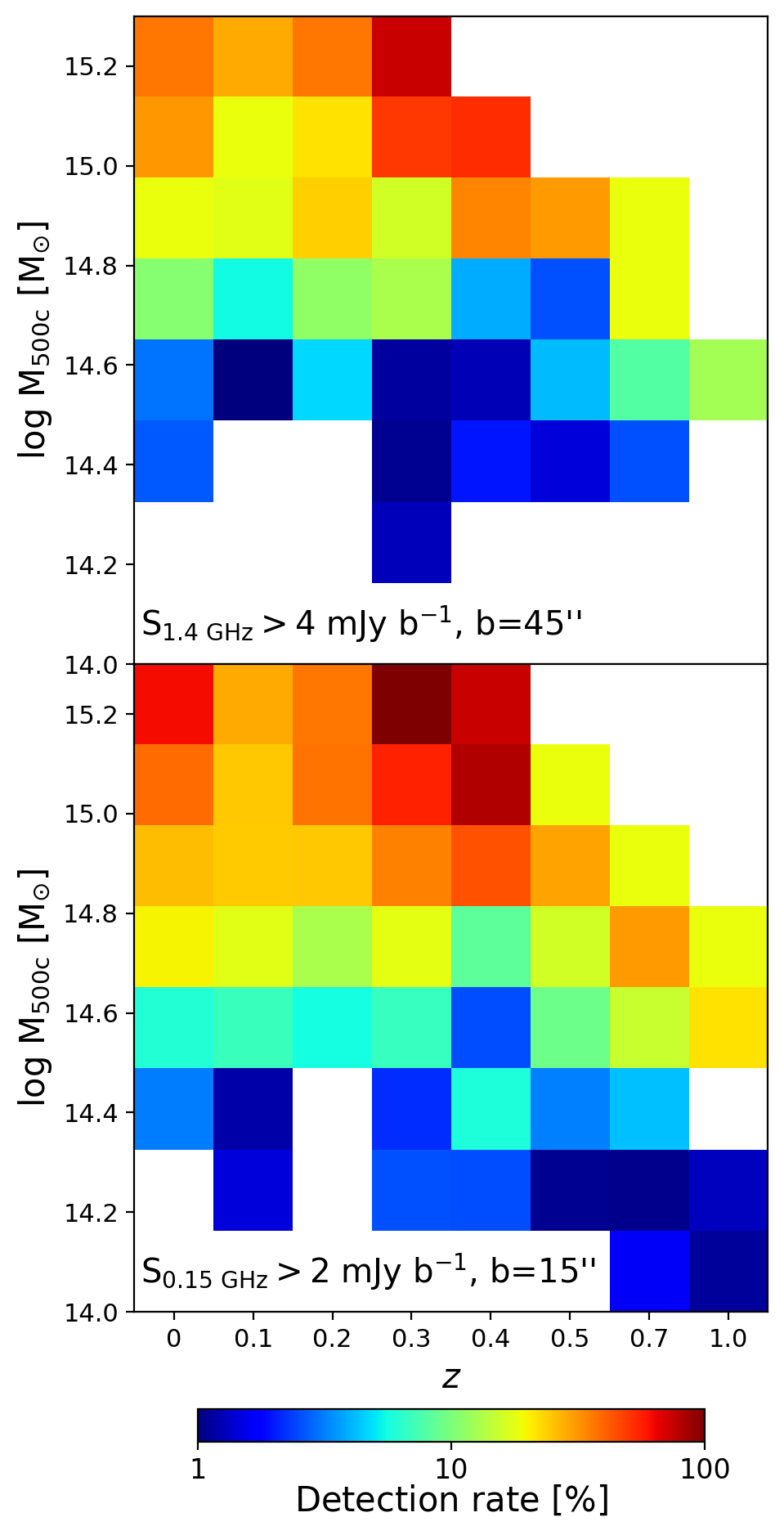}
 \caption{Predicted detection rate of TNG-Cluster radio relics in radio surveys. The radio relics are assumed to be detected when the average surface brightness exceeds $4\rm~mJy~b^{-1}$ in $45''$ beam at $1.4\rm~GHz$ \citep[top, NVSS,][]{2017MNRAS.470..240N} and $2\rm~mJy~b^{-1}$ in $15''$ beam at $0.15\rm~GHz$ \citep[bottom panel, LoTSS,][]{2017A&A...598A.104S}.}
 \label{fig: detection_rate}
\end{figure}

%\subsubsection{Notions on observability and comparison to observations}

The chance to detect radio relics must be lower than the intrinsic occurrence rate quantified above, as observations are limited by sensitivity and classification confusion. 
For a simplistic, zeroth-order comparison with the results of current and future radio surveys, we estimate the average surface brightness of the relics by dividing the radio luminosity by their area. 
We estimate the area as $A\sim LLS^2 \cos \theta$, where $\theta$ is the viewing angle, and assume that the area should be greater than $LLS\times 0.2\rm~Mpc$. 
For each radio relic system, we uniformly sample the viewing angle in $[0,90] \deg$ and consider the relic to be detected if its average surface brightness exceeds the required significance with the observation sensitivity. The radio relic detection rate is defined as the chance to detect the relic at the sampled viewing angles.

Figure~\ref{fig: detection_rate} presents the estimated detection rate of TNG-Cluster radio relics in different radio surveys.
Assuming that radio relics can be detected from the NVSS when the average surface brightness exceeds $\geq4\rm~mJy~b^{-1}$ in the $45''$ beam \citep{2017MNRAS.470..240N}, we expect NVSS to detect radio relics in $\mytilde5\pm2\%$ for $M_{\rm 500c} > 10^{14}~\MSUN$ clusters and $\mytilde18\pm4\%$ for $M_{\rm 500c} > 5\times10^{14}~\MSUN$ clusters. 
With the sensitivity of LoTSS, which will detect radio features with average surface brightness $\gtrsim2\rm~mJy~b^{-1}$ in $15''$ beam at $0.15\rm~GHz$, the detection rate increases to $\mytilde9\pm2\%$ for $M_{\rm 500c} > 10^{14}~\MSUN$ clusters and $\mytilde27\pm4\%$ for $M_{\rm 500c} > 5\times10^{14}~\MSUN$ clusters.
Similarly to the intrinsic occurrence rate shown in Figure~\ref{fig: fraction_mass}, the detection rate increases with cluster mass and remains nearly constant with redshift. 

The detection rate can be converted to a number of radio relics by applying a cluster mass-redshift distribution. For example, we can refer to the mass-redshift distribution of the 220 clusters from the Planck 2nd Sunyaev-Zeldovich source catalog \citep{2016A&A...594A..27P}, which has been used in the recent radio survey \citep[LoTSS, ][]{2022A&A...660A..78B,2023A&A...680A..31J}. 
With the TNG-Cluster result and the applied observability criteria, we predict $\mytilde34$ radio relics to be detected (i.e., a detection rate of $15\%$). Within the uncertainties, this is consistent with the observed detection rate of $12\pm7\%$ from LoTSS from the Planck cluster sample \citep{2023A&A...680A..31J}. 

Although these abundance comparisons require further study, we notice a discrepancy between the predicted detection rates of radio relics from this TNG-Cluster study and current radio observations for high-redshift clusters. 
Among the 46 systems at $z>0.5$, our simulation predicts $\mytilde10$ clusters to host radio relics, while so far only two radio relic systems have been reported from LoTSS-DR2 \citep[][]{2023A&A...680A..31J}. Several factors need to be considered to reconcile the difference, including the adopted acceleration efficiency model (Section~\ref{sec: emissivity}), complexity in radio-relic classification or detection, and selection function of observed clusters.

Regarding modeling uncertainties, our emission calculations assume a fixed fraction of shock-dissipated energy converts into nonthermal emission. 
The discrepancy may suggest that the high-redshift cluster environment instead has a lower acceleration efficiency than the one we have assumed in our post-processing. On the other hand, the mass-weighted average Mach numbers of the simulated radio relics are constant with redshift (Section~\ref{sec: relic_stat}).

Misclassification of high-redshift radio features may be another factor for the discrepancy.
Our simulated radio features are considered radio relics whenever their average surface brightness exceeds a certain sensitivity threshold. In contrast, observed radio relics are typically defined as radio features that are spatially separated and distinct from cluster X-ray emission. As this condition cannot be met when the merger occurs along the line of sight, the number of observed radio relics may be lower than reality (see further and related discussion in Section~\ref{sec: halo}).

Finally, we note that there are high-redshift radio relic systems, such as ``El Gordo'' \citep[$z=0.87$,][]{2014ApJ...786...49L} and PSZ2 G091.83+26.11 \citep[$z=0.82$,][]{2023A&A...675A..51D}, which are not included in the samples of \citet{2023A&A...680A..31J}. Thus, it is possible that the observed abundance of high-redshift radio relics might change for the different samples of clusters. 
As a reference, we predict $\mytilde154$ radio relics with $\mytilde22$ systems at redshift $z>0.5$ from the entire 1059 Planck-SZ clusters with $M_{\rm500,~SZ} > 10^{14}~\MSUN$.
Future observational studies with larger numbers of clusters and a direct comparison with simulation results using mock observations are necessary.

\subsection{Scaling relations of radio relic properties} 
\label{sec: relic_stat}

To date, the formation mechanism of extremely large ($\gtrsim 2\rm~Mpc$) radio relics has posted a challenge for cosmological simulations.
In Figures~\ref{fig: Halo3_radio} and ~\ref{fig: reliczoo}, we show that TNG-Cluster naturally reproduces giant radio relics. We now quantify this finding in detail. 

\begin{figure}
 \centering
 \includegraphics[width=\columnwidth]{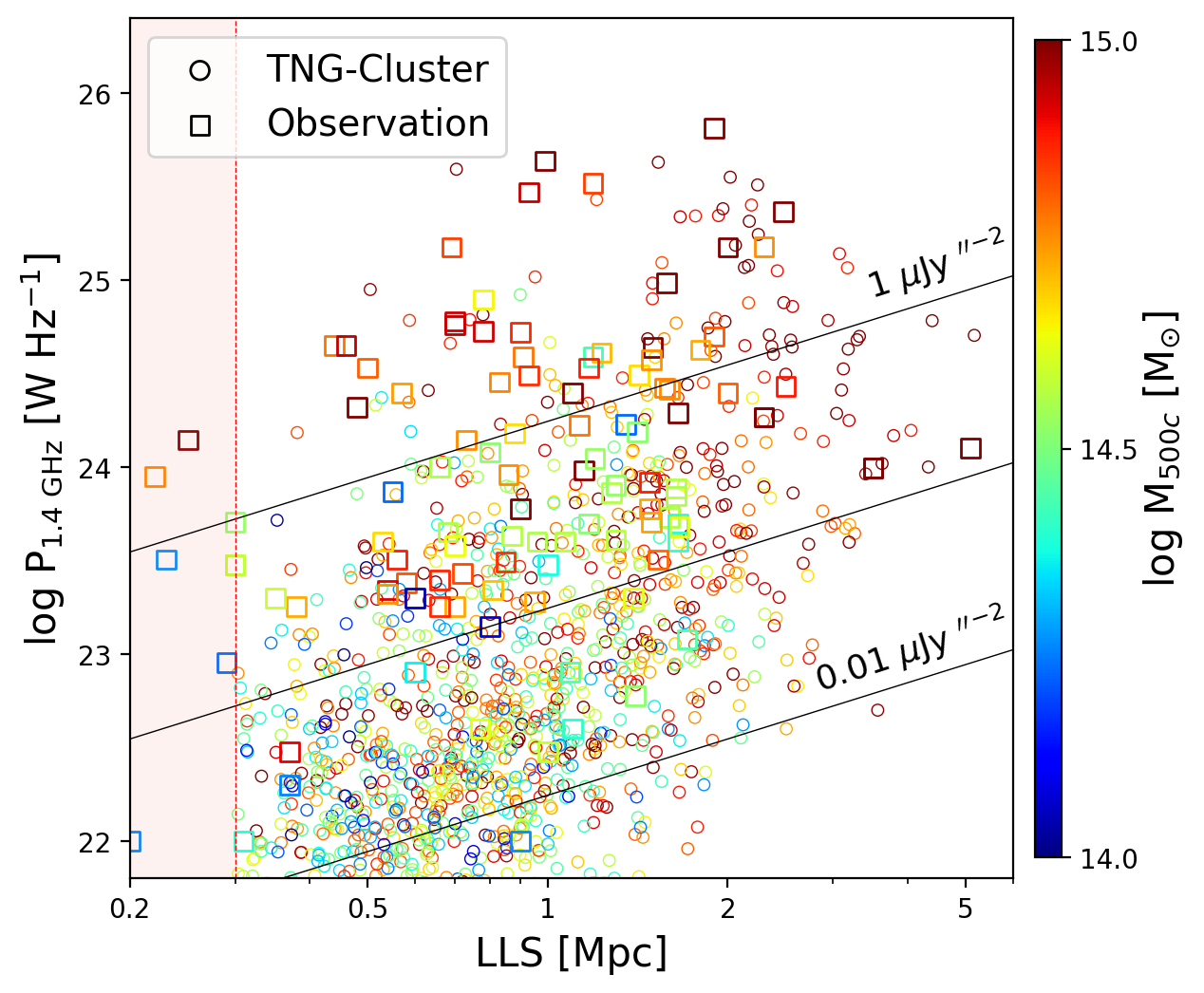}
 \caption{Largest linear size vs. luminosity of TNG-Cluster radio relics. The color encodes $M_{\rm 500c}$ of the host cluster. The largest separation is measured between the cells with $P_{\rm radio}>10^{26}\rm~erg~Hz^{-1}$ to avoid overestimation of radio relic sizes due to the overlapping with accretion shockwaves. 
 Solid lines mark the radio luminosity of the relic with the average surface brightness $1,~0.1,$ and $0.01\rm~\mu Jy~''^{-2}$, assuming the relic redshift $z=0.2$ and width $0.2\rm~Mpc$.
 The dotted line marks the lower limit of the available size of the simulated radio relic.
 The majority of the simulated radio relics larger than $>2\rm~Mpc$ are identified in massive merging clusters with $M_{\rm 500c}>10^{15}~\MSUN$.}
 \label{fig: LLS-relation}
\end{figure}

Figure~\ref{fig: LLS-relation} illustrates that the TNG-Cluster simulation suite yields a number of radio relics with $LLS>2\rm~Mpc$, with the largest one up to $\mytilde 5\rm~Mpc$, hence encompassing the size of the largest radio relic ever observed ($\mytilde3.6\rm~Mpc$) and its tentative upper limit \citep[$\mytilde5.1\rm~Mpc$,][]{2021A&A...656A.154H}.
We find that radio relics of size $>2\rm~Mpc$ are typically created by massive cluster-cluster mergers with a median mass of the system of $M_{\rm 500c}\sim8\times10^{14}~\MSUN$. This highlights the importance of simulating truly massive clusters using a large (Gpc) cosmological box to reproduce and interpret radio relics beyond the Mpc-size. 
In certain instances, lower-mass clusters ($M_{\rm 500c}\sim 10^{14}\rm~\MSUN$) can host Mpc-size radio relics, consistent with existing observational examples \citep[e.g., Abell 168,][]{2018MNRAS.477..957D}.  

The size and luminosity of the simulated radio relics show a weak correlation (Figure~\ref{fig: LLS-relation}). 
However, we caution that this apparent relation is affected by our radio-relic identification scheme.
As detailed above, we measure the size of the simulated radio relics after filtering out cells with a luminosity below $10^{24}\rm~erg~Hz^{-1}$, which inevitably creates a lower limit in luminosity. 
Assuming $200\rm~kpc$ thickness on the $2\rm~Mpc$-size relic, the minimum luminosity becomes $\mytilde10^{22}\rm~W~Hz^{-1}$, which is comparable to the lower limit presented in Figure~\ref{fig: LLS-relation}.
Similarly, observed radio relics suffer from a selection effect, with the majority of detected radio relics having a high surface brightness \citep[see also ][]{2017MNRAS.470..240N}. 
Recent low-frequency radio observations have started to report faint radio relics, with $10^{22}\rm~W~Hz^{-1}$ \citep[e.g.,][]{2022ApJ...925...91S}. Since our simulation predicts a large number of faint radio relics, this suggests that many new faint radio relics may be discovered in upcoming radio surveys \citep[e.g.,][]{2020MNRAS.493.2306B}.

\begin{figure}
 \centering
 \includegraphics[width=0.9\columnwidth]{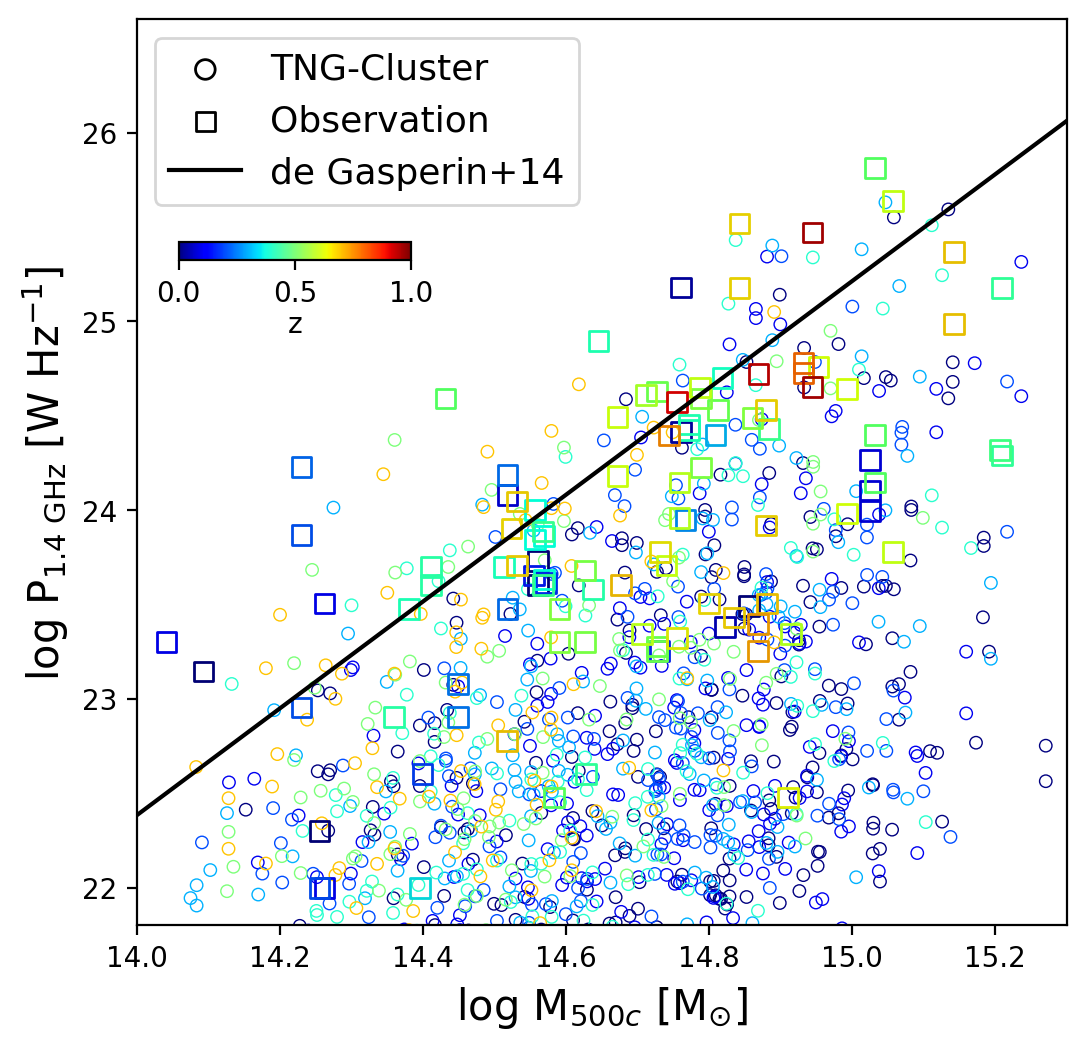}
 \caption{Mass-luminosity relation of TNG-Cluster radio relics (circles) and observed relics (squares). The color denotes redshift. The line is the mass-luminosity relation by \citet{2014MNRAS.444.3130D}. The luminosity of the brightest radio relics increases with cluster mass, while there is a large abundance of low luminosity radio relics, that can be detected by future surveys.
 }
 \label{fig: MassLumRelation}
\end{figure}

In Figure~\ref{fig: MassLumRelation}, we quantify the relationship between radio relic luminosity and host halo mass mentioned earlier (Section~\ref{sec: relic_count}). 
According to TNG-Cluster, as expected, the luminosity of the brightest radio relics in each mass bin increases with cluster mass. 
For any given halo mass, the brightest radio relics roughly follow the observed relation of radio relics (solid black line). Therefore, as suggested in previous studies, this implies that the cluster mass itself does not dictate the luminosity of radio relics, but rather sets an upper limit on possible luminosities \citep[e.g.,][]{2017MNRAS.470..240N,2023A&A...680A..31J}.

By performing a least-square fitting of a power-law relation $P_{\rm 1.4~GHz} = P_{\rm0,~1.4~GHz}\times (M_{\rm 500c}/10^{14} \MSUN)^{\alpha}$, we find that, for the entire TNG-Cluster sample with $M_{\rm 500c}>10^{14}~\MSUN$, the radio relic luminosity increases with cluster mass following the slope and normalization of $\alpha\sim1.5\pm0.1$ and $P_{\rm0,~1.4~GHz}\sim5\times10^{21}\rm~W~Hz^{-1}$, respectively.
However, as discussed earlier, the number of low-mass clusters is small in the TNG-Cluster compared to a volume-limited sample, which causes a Malmquist bias in the mass-luminosity relation. We correct for the Malmquist bias by weighting the cluster with the number ratio of the simulated and theoretical mass functions at the given mass bin.
To derive the theoretical mass function, we use the {\tt Colossus} package \citep{2018ApJS..239...35D} and the halo mass function of \citet{2008ApJ...688..709T}.
After correction, the slope of the fitted power law becomes steeper ($\alpha\sim2.0\pm0.1$), while the normalization remains consistent with the pre-correction value.

The slope of the mass-luminosity relation of TNG-Cluster radio relics is comparable to theoretical expectations.  
The energy injected from the merger is proportional to $E\propto M^2/R$, and the power of the radio relic can be described with $P\propto E/t$, where $t$ is the lifetime of the radio relics. 
If we assume that the lifetime of the radio relic is proportional to the sound crossing time $t\propto R/c_{\rm s}$ \citep[e.g.,][]{2014MNRAS.444.3130D} and the virial radius $R\propto M^{1/3}$, then the virial equilibrium $T\propto M^{2/3}$ predicts the power to be a function of $P\propto M^{5/3}$, which roughly agrees with our slope. 

On the other hand, the slope of the mass-luminosity relation in TNG-Cluster is shallower than the observed mass-luminosity relation $P\propto M^{2.8}$ \citep{2014MNRAS.444.3130D}. We believe that this is simply due to the selection bias in the observed samples. In observations, the number of clusters per area increases with redshift as the volume increases, whereas only the brightest radio relics are likely to be detected in the high-redshift regime.
Therefore, this will preferentially select the bright radio relics at the high mass end, which will steepen the slope of the relation. 

With TNG-Cluster, we find that the normalization of the mass-luminosity relation increases with redshift (see color coding of Figure~\ref{fig: MassLumRelation})
For instance, if we fit the mass-luminosity relation with the fixed slope of $P \propto M^{2.0}$, the normalization steadily increases with redshift from $10^{21.5}\rm~W~Hz^{-1}$ at $z=0.0$ to $10^{22.2}\rm~W~Hz^{-1}$ at $z=1.0$. 
Moreover, the brightest radio relics of low-mass clusters are often at high redshift. The brightest radio relic with $M_{\rm 500c}<~10^{14.6}~\MSUN$ occurs in TNG cluster at $z=1.0$ with $P_{\rm 1.4~GHz}\sim8\times10^{24}\rm~W~Hz^{-1}$ (panel {\tt p} in Figure \ref{fig: mw_relic_zoo}). This is more than an order of magnitude brighter than the brightest radio relic in the same mass range at $z=0.0$ ($P_{\rm 1.4~GHz}\sim3\times10^{23}\rm~W~Hz^{-1}$).
A similar trend in normalization has been reported in \citet{2011ApJ...735...96S}. 

\begin{figure}
 \centering
 \includegraphics[width=\columnwidth]{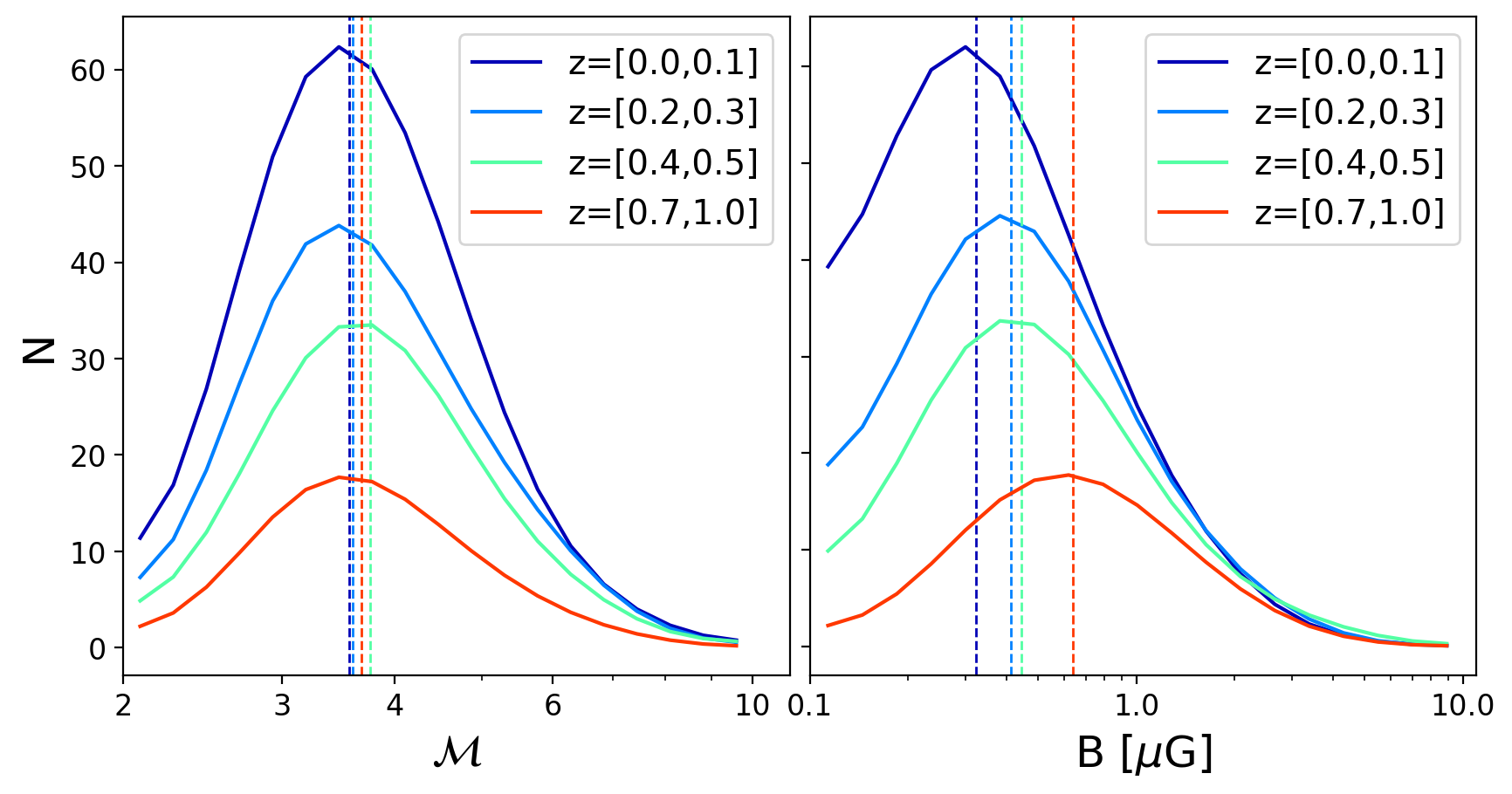}
 \caption{Distribution of mass-weighted average Mach number (left) and magnetic field magnitude (right) of TNG-Cluster radio relics at different redshifts. The average Mach number of the simulated radio relics is constant in time, with $\mathcal{M}\sim4$. In contrast, the average magnetic field strength in the regions of the radio relics increases by a factor of $\mytilde2$ between $z=0$ and $z=1$.}
 \label{fig: Mach_n_Bfield}
\end{figure}

We speculate that the redshift evolution in the normalization of the radio-relic mass-luminosity relation naturally emerges because of the higher magnetic field magnitude at higher redshift \citep[see also ][]{2023arXiv231106338N}. 
Figure~\ref{fig: Mach_n_Bfield} shows the mass-weighted average Mach number and the magnetic field strength distributions of TNG-Cluster radio relics.
Overall, typical radio relics are generated by weak shockwaves with $\mathcal{M}\sim4$ and with a magnetic field strength of $B\sim0.5\rm~\mu G$, in qualitative agreement with the inferred physical properties of observed radio relics (see, \citet{2019SSRv..215...16V} and references therein). Furthermore, the median Mach number of TNG-Cluster radio relics is constant across the redshift range. 
On the other hand, the median magnetic field strength nearly doubles as the redshift increases from $z=0.0$ to $z=1.0$.
This evolutionary factor increases to $\mytilde2.4$ if we fix the cluster mass range to $M_{\rm 500c}=[10^{14},10^{14.7}]~\MSUN$; a sufficient number of clusters exist up to $z=1$ in this mass interval.
As the radio luminosity is proportional to the magnetic field strength with $\propto B^{1+(s/2)}$, the evolution of the simulated magnetic field strength is expected to amplify the normalization in the mass-luminosity relation by a factor of $\mytilde5$, which agrees with our fitting results. 
In turn, the larger magnetic fields at high redshifts can be explained by the denser environment in cluster outskirts where radio relics develop. This trend can also explain the increasing occurrence rate of radio relics, although we note that the number of massive clusters at high redshift is smaller.

\section{Discussion}
\label{sec:discussion}

\subsection{Origin of the diverse radio relic luminosities}
\label{sec: diverse}

The radio relics simulated and discussed in this paper have a large scatter in their mass-luminosity relation, including very low-luminosity relics in high-mass merging clusters, as shown in Figure~\ref{fig: MassLumRelation}. This suggests that additional parameters and physical factors may determine their final luminosity. 
In the following, we check and discuss plausible secondary parameters, in addition to the mass of merging systems, that can justify the relic-to-relic variation predicted by TNG-Cluster. 

Firstly, we emphasize that although the relic luminosity is dependent on the model parameters in post-processing (see previous Sections), the system-to-system variation is not sensitive to the choices.
Also, to minimize the impact of the cluster mass, we normalize the relics' length properties with $R_{\rm 500c}$ and the radio luminosity using the mass-luminosity relation $P_{\rm fit}=5\times10^{21}\rm~W~Hz^{-1}\times (M_{\rm 500c}/10^{14}\rm~\MSUN)^{2.0}$ (Section~\ref{sec: relic_stat}). 
Henceforth, we refer to the ratio $P_{\rm 1.4~GHz}/P_{\rm fit}$, as normalized luminosity. 

\begin{figure}
 \centering
 \includegraphics[width=\columnwidth]{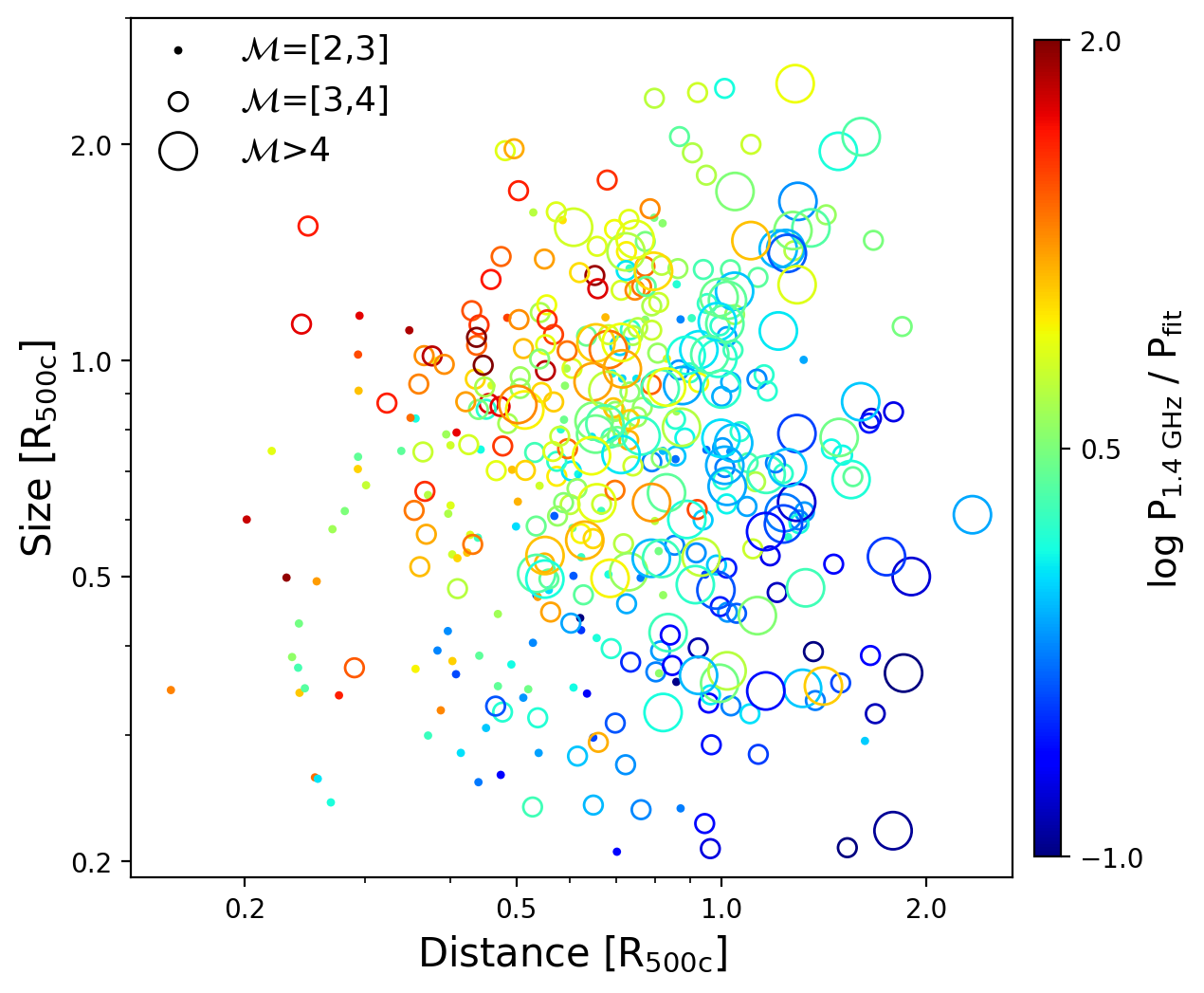}
 \caption{Normalized distance from the cluster center vs. largest linear size of TNG-Cluster radio relics. The color depicts the luminosity of each relic normalized with the luminosity derived from the mass-luminosity relation. The size marks the mass-weighted Mach number of the relic. Radio relics at smaller clustercentric distances and with larger physical sizes are brighter.}
 \label{fig: relic_dist}
\end{figure}

One possible source of the scatter may be connected to the phase (i.e., the evolutionary stage) of the radio relics.
Studies have claimed that the radio relic luminosity evolves rapidly in time after the collision \citep[e.g.,][]{2011ApJ...735...96S,2012MNRAS.421.1868V}. According to such studies, the luminosity of the relic is small shortly after the collision, as shock compression is hindered near the cluster center. As the radio relic propagates to the outskirts of the cluster, its luminosity increases and peaks at around $\sim$1~Mpc separation from the center of the cluster, and then decreases as the shock of merger reaches the periphery of the ambient cluster. 
In Figure~\ref{fig: relic_dist}, we compare the clustercentric distance of the radio relics and their sizes, color-coded with the normalized luminosity.
The normalized luminosity of relics at a given size can vary up to three orders of magnitude, depending on their distance to the cluster center. In fact, according to TNG-Cluster, radio relics closer to the center are brighter than those positioned beyond $R_{\rm 500c}$, while the number of radio relics drops at a small cluster-centric distance ($\lesssim0.2R_{200\rm c}$). 
Therefore, even though the picture emerging from the TNG-Cluster simulation is not aligned with the aforementioned previous theoretical reasoning, Figure~\ref{fig: relic_dist} clearly shows that radio relics inspected in any given simulation snapshot constitute a random selection of their evolutionary phase, hence producing large scatter in the mass-luminosity relation.

However, the existence of a number of bright TNG-Cluster radio relics at $\lesssim0.5R_{200\rm c}$ is in qualitative conflict with observations, since few radio relics have been reported in the proximity of the cluster centers \citep[][]{2019SSRv..215...16V}.
We can consider a few possibilities that might explain such a discrepancy. 
First, our acceleration model, which assumes a fixed fraction of shock-dissipated energy, might overpredict bright radio relics close to the cluster center \citep[e.g.,][]{2012MNRAS.421.1868V}. 
Nevertheless, the impact might be negligible, as the strong shockwave within the broad distribution of the shock strength still dominates the radio emission. The median of the radio emissivity-weighted average Mach number of these weak shocks is $\mathcal{M}_{\rm radio}\sim3.8$, which is large enough to generate magnetic instabilities to drive the diffusive shock acceleration. 
Second, the difference in the scheme used to classify and hence identify radio relics between observations and simulations can also lead to discrepancies, as previously discussed in the context of the abundance of radio relics at high redshifts (Section~\ref{sec: relic_stat});  radio relics are often defined by their position relative to the X-ray emission, and the classification becomes unclear when the relic lies close to the cluster center \citep[e.g.,][]{2020ApJ...897...93B}. 
Finally, different definitions of the cluster center (e.g., based on mass, X-ray, or galaxies) may alter the distance measurements. 

\begin{figure}
 \centering
 \includegraphics[width=0.9\columnwidth]{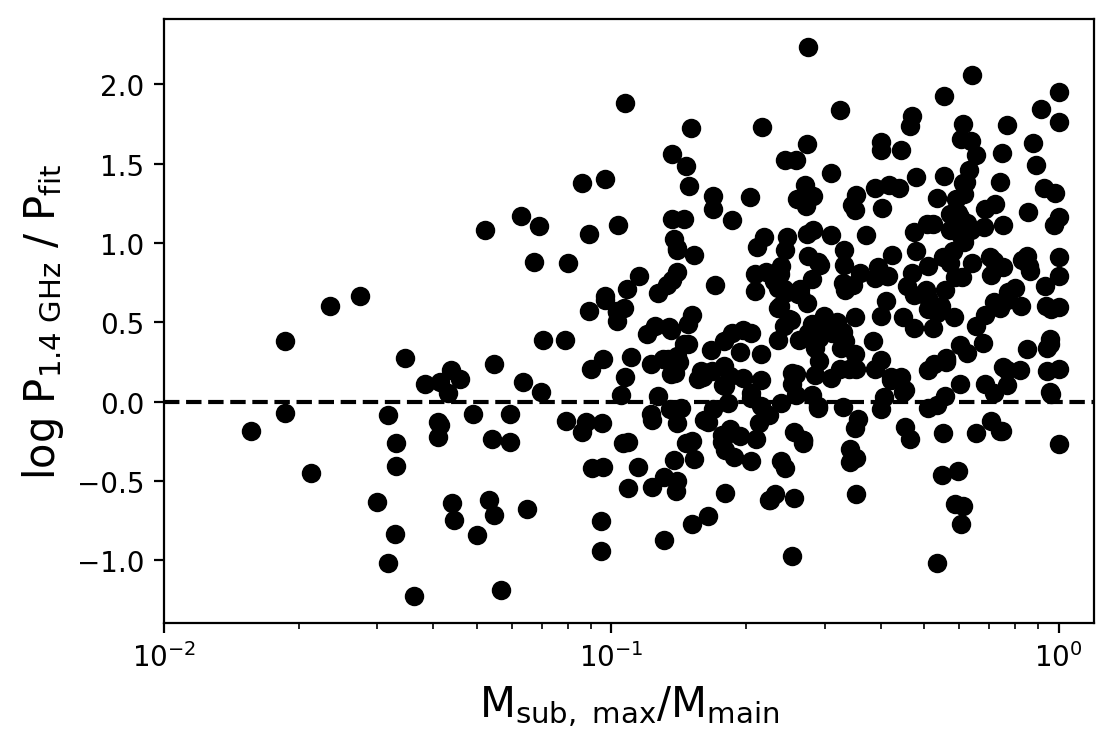}
 \caption{Mass ratio vs. normalized radio luminosity of TNG-Cluster radio relics. The luminosity is normalized with that derived from the mass-luminosity relation. The mass ratio is defined using the most massive subcluster that experiences a collision with the main cluster.}
 \label{fig: massratio}
\end{figure}

Going back to the origin of the scatter in the mass-luminosity relation, the mass ratio of the cluster merger may be another plausible culprit. 
In Figure~\ref{fig: massratio}, we show the distribution of the mass ratios and the normalized relic luminosity.
Here, we use the mass ratio with the most massive merging subcluster (i.e., \textsc{SUBFIND} subhalo), as the primary cluster can experience multiple merger events in a short period of time. Despite the large scatter, we can notice a clear trend: radio relics are likely to be brighter, relative to the average mass-luminosity relation, if caused by a major rather than a minor merger. 
As discussed in the context of Figure \ref{fig: MassLumRelation}, observations preferentially select brighter radio relics at a fixed cluster mass. Therefore, we expect that the majority of observed systems are major mergers, which is consistent with observational constraints \citep[e.g.,][]{2001ApJ...553L..15B,2022AAS...24021405F}. 

\begin{figure}
\centering
\includegraphics[width=\columnwidth]{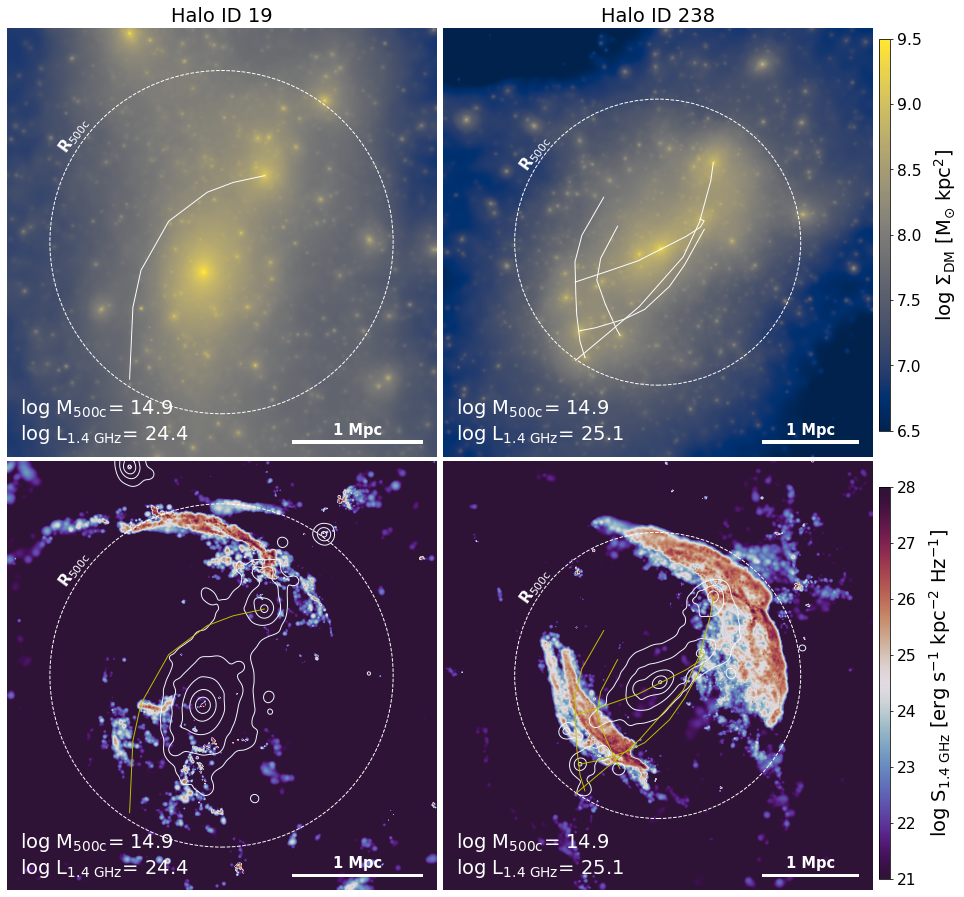}
 \caption{Dark matter surface density (top) and radio surface brightness map (bottom row) of two TNG-Cluster mergers with comparable merger properties, including halo mass ($\log M_{\rm 500c}\sim14.9$), mass ratio ($\mytilde4$), and time since collision ($\mytilde0.4\rm~Gyr$). The overlaid curves mark the orbit of the merging subcluster for the last $\mytilde1\rm~Gyr$, which has $M_{\rm sub,~IC}>5\times10^{12}~\MSUN$ and experienced the first pericenter passage in the $\mytilde1\rm~Gyr$. The radio relics produced in the two systems are different due to the specific details of the merger history. 
 }
 \label{fig: mult_merger}
\end{figure}

Finally, thanks to the wealth of the TNG-Cluster relic sample and thus the large distribution in their physical properties, we can compare two cluster merger cases with a comparable halo mass, mass ratio, and evolutionary phase, but with very different radio relic configurations.
Figure~\ref{fig: mult_merger} presents such an example. 
Here, the two merging clusters are both massive ($\log M_{\rm 500c}\sim14.9$) and experienced a recent ($\mytilde0.4\rm~Gyr$) major merger with a mass ratio of $\mytilde4$. While the merger conditions are similar, the two radio relic systems are very different. Halo 238 presents bright double radio relics with $\log P_{1.4\rm~GHz} =25.1$. On the contrary, Halo 19 shows a single relic with a relatively low radio power $\log P_{1.4\rm~GHz} =24.4$. 

\begin{figure*}
\centering
\includegraphics[width=2\columnwidth]{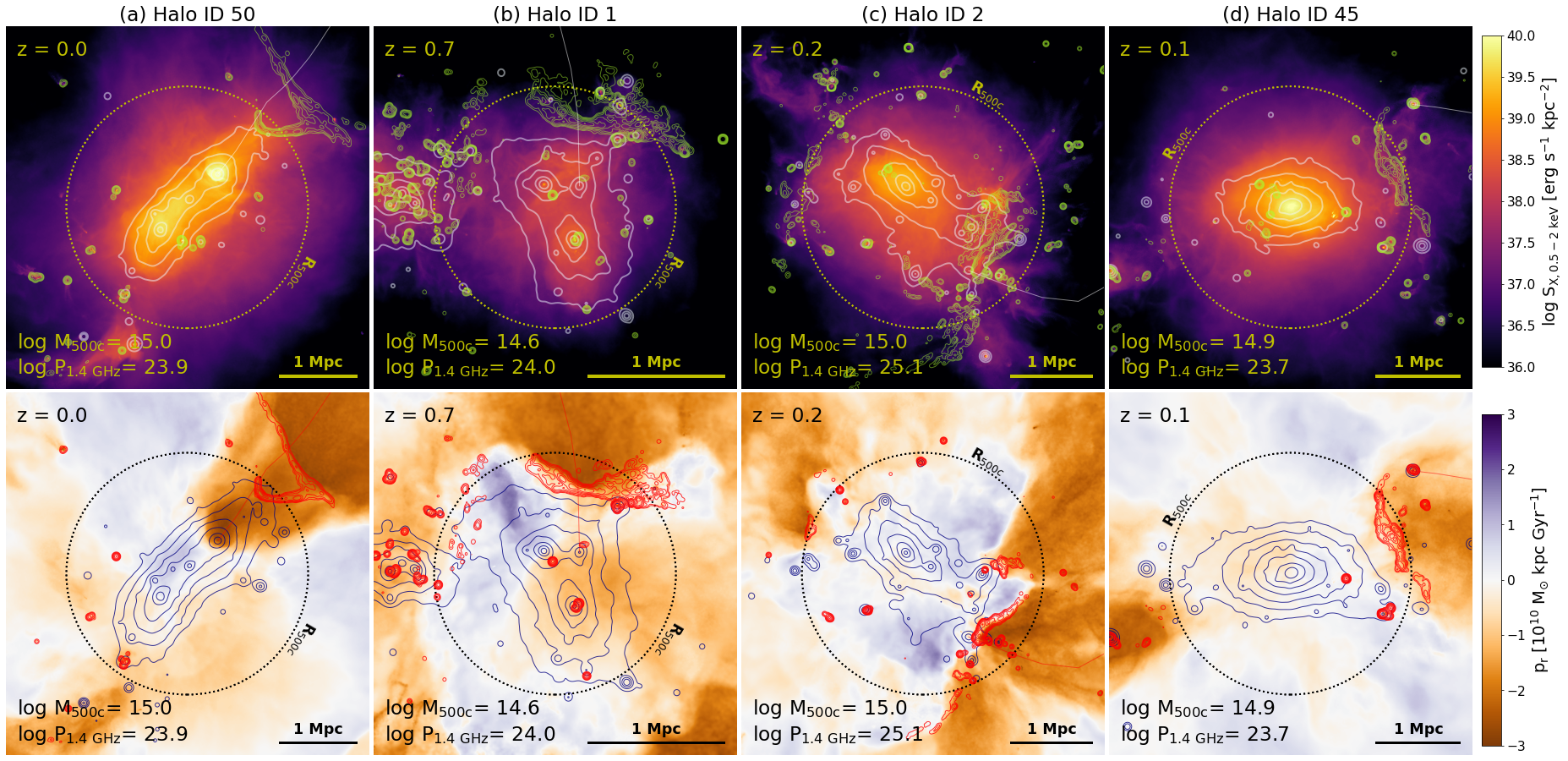}
 \caption{Examples of inverted radio relic systems from TNG-Cluster. On the top X-ray surface brightness maps are overlaid with the mass (white) and radio contours (green). The bottom row provides $500$-kpc slices of the radial momentum, whereby the midplane and center of the slice pass through the relic and cluster centers. The dotted circle marks the $R_{\rm 500c}$ of the cluster. The inverted radio relics coincide with an infalling motion of the gas. The solid line marks the path of the infalling subhalo, which we suspect creates the inverted morphology of the relics.
 }
 \label{fig: inverted}
\end{figure*}

%What determined the significant differences in radio relics for the two systems shown in Figure~\ref{fig: mult_merger}? 
The significant difference in radio relics for the two systems shown in Figure~\ref{fig: mult_merger} is likely due to differences between the merger histories of the two clusters. 
In the case of Halo 19, a single subcluster with $1.2\times10^{14}\rm~\MSUN$ mass has collided with the main cluster, which generates a single radio relic in front of its path. 
On the other hand, Halo 238 has experienced a collision with a cluster of similar mass ($9\times10^{13}\rm~\MSUN$), but additionally, two group-scale halos ($\mytilde10^{13}\rm~\MSUN$) and two cluster-scale halos ($\mytilde10^{14}\rm~\MSUN$) have collided along the same collision axis within the last $1\rm~Gyr$. 
These subhalos are massive enough to generate merger shocks, and the radio emissions from the multiple shockwaves superpose to create bright radio relics. 
Thus, we believe that the radio relics in Halo 238 are brighter than those in Halo 19 because of the simultaneous energy injection from the multiple collision events. 

It should be noted that Halo 238 presents arc-shaped double radio relics even after multiple collision events.
We speculate that this is due to the preferred infall direction of massive subclusters along the large-scale structure, which aligns with the merger shock normals. 
Nevertheless, the radio relics formed under this scenario may present a discontinuous shock front or different Faraday depths due to the different positions of the merger shocks in the ICM.

Finally, we note that simultaneous multiple collisions of clusters with other massive systems are common. 
We record the time of the first closest passage of all subgroups and subclusters with total mass $>10^{13}\rm~\MSUN$ in the 95 cluster mergers that host radio relics with $P_{\rm 1.4~GHz}>10^{24}\rm~W~Hz^{-1}$.
Within $\pm0.5\rm~Gyr$ since the collision with the most massive subcluster, we find that $\mytilde76\%$ of the systems have also experienced another collision with a group-scale halo.
As the crossing time of the merger shock is on the order of Gyr, this implies that multiple merger shocks from independent collisions can encounter one another.
This, in turn, can increase the final radio emission due to the superposition of shockwaves, as in Halo 238.
Conversely, the luminosity can decrease if the earlier minor merger preheats the ICM, hindering the formation of a strong merger shock.
We note that, according to the multiple-shock scenario, the successive passage of merger shocks can significantly boost the acceleration efficiency and radio luminosity \citep{2021JKAS...54..103K,2022MNRAS.509.1160I}, which requires further studies to quantify the impact of multiple shocks.  

In summary, the luminosity of the radio relics presents diverse properties, as multiple parameters govern their formation and evolution. 
The potential energy of the primary cluster mass defines the total energy budget of the radio relics.
Then, the radio luminosity can vary, depending on the phase of the radio relic formation, the mass ratio, and the presence of other collisions.

\subsection{Inverted radio relics and their formation} 
\label{sec: Inverted}

Observations have reported a few cases of inverted radio relics that exhibit convex features toward the cluster center, instead of away from it. 
Recently, \citet{2023ApJ...957L..16B} used a zoom-in simulation of a galaxy cluster and proposed that compression of the merger shockwaves by an infalling tertiary subcluster can deform the merger shock and produce inverted radio relics. 

In TNG-Cluster we find a number of inverted radio relics, and four examples are presented in Figure~\ref{fig: inverted}. 
We define the radial momentum as $p_{\rm r}=m\Vec{v}\cdot \hat{r}$, where $\hat{r}$ is a relative position vector and $\Vec{v}$ is the relative velocity of the gas cells with respect to the bulk motion of the primary cluster, according to the halo catalogs.
We note that the systems shown in Figure~\ref{fig: inverted} are centered on the center of the mass of the mergers. 

Similarly to \citet{2023ApJ...957L..16B}, we can identify an infalling subcluster and an infalling gas motion in the direction of the inverted radio relics. 
For example, Halo 50 in panel~{\tt a} is the result of a triple merger.
Here, two main clusters with $4.7$ and $1.8\times10^{14}\rm~\MSUN$ have experienced their first closest passage $\mytilde 2.1\rm~Gyr$ ago, and a third subcluster with $1.6\times10^{14}\rm~\MSUN$ is falling from the top right corner of the image (see the solid line).
As the direction of the subcluster infall is aligned with the collision axis of the post-merger, the gas inflow along the infalling cluster warps the merger shock and produces an inverted radio relic. 
Similarly, Halo 1 in panel~{\tt b} presents another example of a triple merger at high redshift ($z=0.7$). 

These cases exhibit inverted radio relics at early phases of a triple merger, namely when the third pre-merging system compresses the shockwaves produced by the preceding cluster-cluster merger.
According to this scenario, we can expect the inverted radio relic cluster to present an extended X-ray morphology toward the radio relic, such as Abell 1697 \citep{2021A&A...651A.115V}, or three mass peaks that were involved in the collision, as in SPT-CL J2023-5535 \citep{2020ApJ...900..127H}.

The inverted radio relic can become much brighter than its double radio relic pair. 
This is due to the large mass inflow along the third subcluster, which enhances the kinetic flow of the merger shocks. 
Moreover, the shock strength will also increase by the combination of post-merger and pre-merger shockwaves \citep[e.g.,][]{2020MNRAS.494.4539Z}.
Halo 2 (panel {\tt c}) presents an example of this luminosity hierarchy.
Here, the primary massive cluster of $M_{\rm 500c}\sim10^{15}\rm~\MSUN$ has only experienced a minor merger with a $10^{13}\rm~\MSUN$-scale group. 
Nevertheless, this cluster hosts one of the brightest radio relics with $P_{\rm 1.4~GHz}\sim10^{25}\rm~W~Hz^{-1}$, as another $10^{13}\rm~\MSUN$ infalling subgroup has amplified the radio emission of the previous merger shock. 
Its double radio relic pair, which we can identify with patches of radio emission on the opposite sides of the cluster, presents a faint radio emission ($P_{\rm 1.4~GHz}\sim3\times 10^{22}\rm~W~Hz^{-1}$).
The Coma cluster presents an inverted radio relic, and recently a faint radio relic was discovered on the opposite side of the cluster \citep{2022ApJ...933..218B}. 
This could be a possible example of a faint double radio relic pair, although
this scenario requires further investigation to explain its position far from the cluster center $\mytilde R_{\rm 100c}$. 
If the merger shock encounters a subhalo in the late merger phase, then the inverted radio relic will reside far from the cluster center. The double radio relic pair also becomes too faint to be observed (e.g., panel {\tt a}), which can explain the radio relic in the Ant cluster \citep[][]{2021ApJ...914L..29B}.

\begin{figure*}
\centering
\includegraphics[width=2\columnwidth]{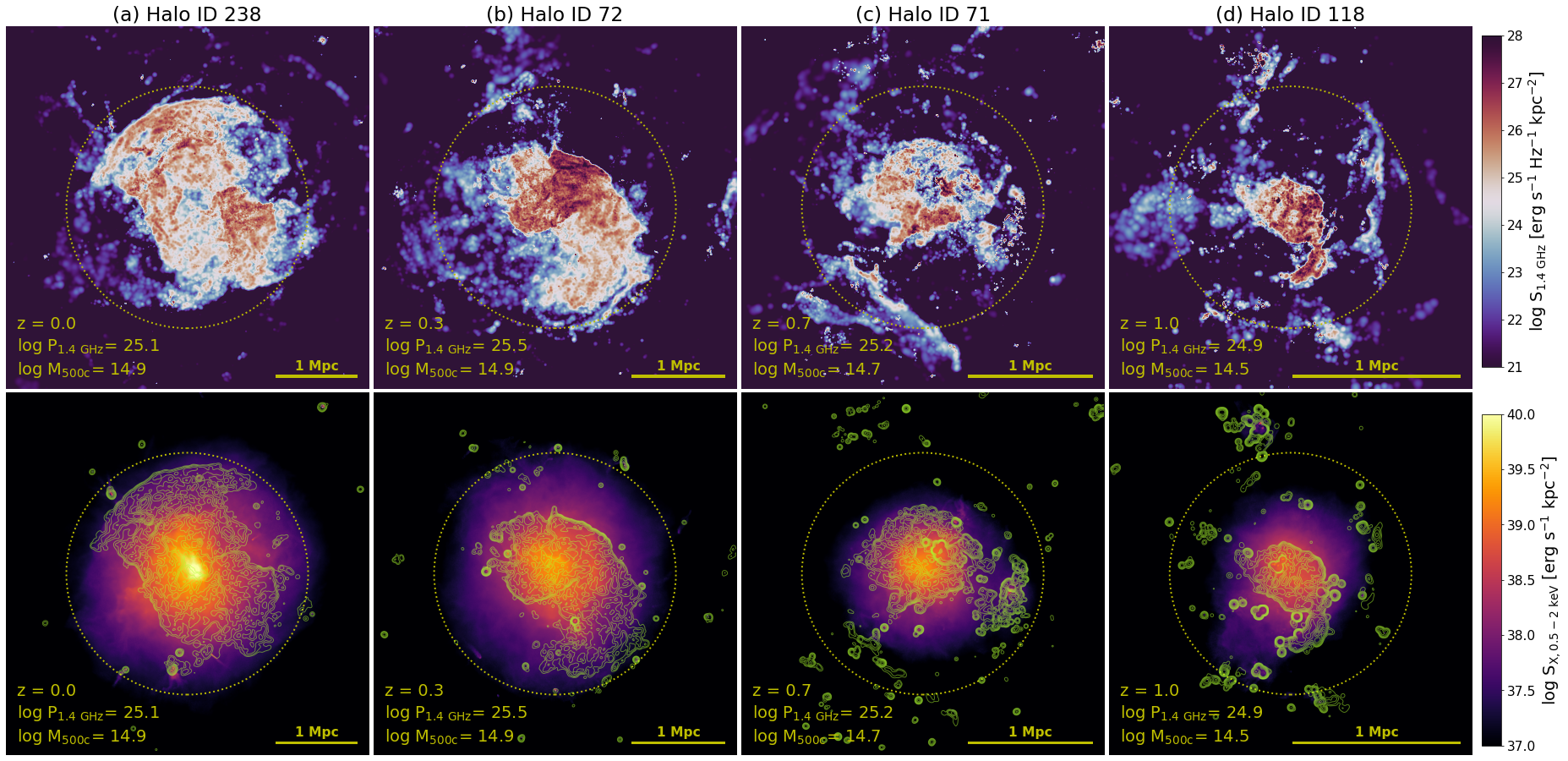}
 \caption{Face-on view of TNG-Cluster radio relic systems that may be confused as radio halos. We plot the radio surface brightness (top) and the X-ray surface brightness map overlaid with the radio contours (bottom). The dotted circle marks the $R_{\rm 500c}$ of the cluster. In these projections, the radio relics exhibit a radio halo-like appearance. However, differently from the latter, they are characterized by a significant drop in surface brightness at their edges and by a radio emission that is not correlated with the X-ray emission. 
 }
 \label{fig: radiohalo}
\end{figure*}

Finally, the case of Halo 45 (panel {\tt d}) presents the formation of an inverted radio relic, where we cannot clearly resolve the two merging clusters in the center.
In this system, we find that the cluster has experienced an equal mass merger event $\mytilde3\rm~Gyrs$ that produced a bright radio relic on the order of $\mytilde10^{25}\rm~W~Hz^{-1}$ at $z\sim0.4$. 
Due to the large mass ratio $M_{\rm sub}/M_{\rm main}$, the merging subclusters have repeatedly collided and generated weak shockwaves until $z=0.1$. 
As a result, when a third subgroup with the mass of $\mytilde10^{13}\rm~\MSUN$ falls into the cluster, the compressed shockwave presents a faint inverted radio relic, whereas the cluster of the central system appears moderately elongated with a relatively relaxed morphology.
We speculate that this scenario can explain the formation of inverted radio relics without multiple mass or X-ray peaks, such as in Abell 3266 \citep{2022MNRAS.515.1871R}.

In summary, we interpret the formation of inverted radio relics as the result of the compression of preexisting merger-induced shocks by a tertiary group-size infall, similarly to what was proposed by \citet{2023ApJ...957L..16B}.
These infalling subgroups might be identified in observations as faint X-ray emissions or bright galaxy groups. However, their unambiguous identification would be challenging, as the spatial separation between the radio relics and the infalling subgroups can vary.
A hint of large-scale structure toward inverted radio relics are an alternative signature of an infalling subgroup, requiring a statistical study on the observed inverted radio relics.

\subsection{Confusion with radio halos} 
\label{sec: halo}

Upcoming radio surveys such as the SKA will detect more faint radio relics in the future. Since the viewing angle of these relics will be random, some of them will be viewed face-on (i.e., propagating along the line-of-sight direction), and their projected radio features will appear similar to those found in "radio halos", as illustrated in Figure~\ref{fig: Halo3_radio}.
They can also overlap with the cluster X-ray emission through projection and produce unpolarized radio emission because the magnetic fields in the relics are mostly aligned parallel to their extent.
Therefore, we need guidelines to avoid misclassifying face-on radio relics as radio halos. 

Figure~\ref{fig: radiohalo} shows four examples of face-on radio relics in TNG-Cluster, which present a morphology similar to radio halos. 
Now, despite the apparent morphological similarities, unlike radio halos whose brightness profile decreases gradually with increasing distance from the cluster center \citep[e.g.,][]{2009A&A...499..679M}, the surface brightness of face-on radio relics truncate abruptly at their boundaries, this being due to the radio emission of the merger shock that vanishes outside the shockwaves \citep{2020A&A...634A..64B}. 
We note that this surface brightness discontinuity is different from the radio halo discontinuity reported in recent radio observations \citep[e.g.,][]{2023A&A...674A..53B}, as the face-on relics do not feature X-ray discontinuities along the boundary. 

The lack of a spatial correlation between the radio and X-ray surface brightness distributions can also help alleviate the confusion.
The center of the face-on relics coincides with the X-ray center as is the case for radio halos. 
However, radio relics trace the merger shock energy distribution. At the same time, the X-ray emission closely traces the density of ICM, which should result in a lack of correlation between the surface brightness distributions of the radio and the X-ray. 
Indeed, when we sample the regions within $R_{\rm 500c}$ and compare the radio and X-ray surface brightness measured with $100\rm~kpc$-size bins above $S_{\rm 1.4~GHz}>0.01\mu Jy~''^{-2}$ and $S_{\rm X-ray}>10^{38}\rm~erg~s^{-1}~kpc^{-2}$, the four examples displayed in Figure~\ref{fig: radiohalo} show a weak correlation ($I_{\rm 1.4~GHz}\sim I_{\rm X}^{0.2}$) compared to those reported in radio halo observations \citep[$I_{\rm 1.4~GHz}\sim I_{\rm X}$,][]{2001A&A...369..441G}. 
 
Observed radio relics may blend with radio halos \citep[e.g.,][]{2018ApJ...852...65R}.
Moreover, the radio features are smoothed by the synthesized beam of the radio observations, making the classification of face-on relics using a surface brightness drop and lack of a spatial correlation difficult, especially for systems at high redshifts, such as Halo 118 in panel {\tt d} of Figure~\ref{fig: radiohalo}.
We speculate that cluster dynamics can help resolve this ambiguity. Radio relics extend perpendicular to the collision axis, so face-on relics are expected when the cluster merger is viewed along the line-of-sight direction. Therefore, multi-object spectroscopy or kinetic SZ observations, which can distinguish the line-of-sight velocity structures \citep[e.g.,][for a review]{Sayers2019,Mroczkowski2019}, can be useful in alleviating the misclassification of relics as halos.

\section{Summary} 
\label{sec:summary}

This paper presents a large sample of simulated radio relics extracted from the TNG-Cluster simulation - a suite of magnetohydrodynamical cosmological zoom-in simulations of 352 massive galaxy clusters with $\MFIVEC = 10^{14.0-15.3}~\MSUN$ sampled from a 1 Gpc-size cosmological box performed using the moving-mesh code AREPO. This enables unprecedented studies of radio relics originating from cluster-cluster and group-cluster mergers in the full cosmological context, including mergers with high-velocity collisions, significant pericenter separations, and rare examples at high redshift. Our main results can be summarized as follows:

\begin{itemize}

    \item We demonstrate that the TNG-Cluster simulation can produce analogs of observed radio relics using a basic post-processing model based on nonthermal plasma acceleration. We can reproduce their abundances (Figure~\ref{fig: detection_rate}), statistical relationships (Figure~\ref{fig: MassLumRelation}), and diverse morphologies (Figure~\ref{fig: reliczoo}), including very extended (> Mpc) relics and multiple analogs to observed systems (Figures~\ref{fig: mw_relic_zoo} and \ref{fig: LLS-relation}). The shocks resulting from the cosmological and hierarchical development of structure in TNG-Cluster can account for the origin of diffuse synchrotron emission in cluster mergers.

    \item In TNG-Cluster, merging clusters have radio relics with a wide range of morphologies and properties, such as double and single relics, linear and arc-shaped systems, and even "inverted" relics (shown in Figure~\ref{fig: reliczoo}). We derive the radio emissivity of shockwaves that are naturally produced in the simulated systems, by applying an analytic model to accelerate cosmic-ray electrons (CRe) to the shock properties identified by an on-the-fly shock finder. This indicates that the diverse morphology of these relics may not be solely due to inhomogeneous acceleration efficiency but rather a result of the complex shocks generated by the cosmological, hierarchical assembly processes.
    
    \item Radio relics can serve as a robust indicator of the collision axis. When merger-driven shocks are launched and propagate along the collision axis, it becomes possible to determine the collision axis by analysing the location, orientation, and pair-arrangement of the radio relics. This remains valid even during a complicated collision, where the spatial extent of the mass and X-ray emission may be misaligned with the collision axis (see Figure~\ref{fig: mw_relic_zoo}). Among the approximately $300$ simulated and identified TNG-Cluster radio relics, many visual analogs to observed cluster mergers and radio relics can be identified. These analogs can be used as a reference to reconstruct their collision scenario.

    \item According to TNG-Cluster and assuming our model for CRe production, radio relics are relatively rare. Of $M_{\rm 500c}>10^{14}~\MSUN$ clusters, only about $14\%$ host radio relics with $P_{\rm 1.4~GHz}>10^{23}\rm~W~Hz^{-1}$ and only 20 clusters contain radio relics with $P_{\rm 1.4~GHz}>10^{25}\rm~W~Hz^{-1}$ in the redshift range $z=0-1$ (Figure~\ref{fig: fraction_mass}).
    The occurrence rate of radio relics with $P_{\rm 1.4~GHz}>10^{23}\rm~W~Hz^{-1}$ increases to $\mytilde38\%$ in more massive clusters with $M_{\rm 500c}>5\times10^{14}~\MSUN$, whereas the rate is constant within the uncertainties at different redshifts. 

    \item The expected number of radio relics is in agreement with that of recent surveys. For example, by applying the cluster mass-redshift distribution of the Planck 2nd SZ source catalog that was used in the LoTSS survey, we find that the probability of detecting radio relic systems is $\mytilde15\%$, consistent with recent observations within the uncertainty ($\mytilde12\pm7\%$, Figure~\ref{fig: detection_rate}). However, the TNG-Cluster simulation seems to predict a larger number of high-redshift radio relics than observations.
    We speculate that the discrepancy may be due to inefficient acceleration, difficulty detecting and classifying radio relics, or cluster selection functions in high-redshift clusters.  
    We estimate that approximately $154\pm 40$ radio relic systems will be detectable at the current LoTSS survey depth, of which roughly $22$ are at $z>0.5$, from $1059$ Planck-SZ clusters with $M_{\rm 500c}>10^{14}~\MSUN$, without accounting for potential misclassifications. 
    
    \item We have discovered that extremely large radio relics with extensions greater than $2\rm~Mpc$ are typically found in massive cluster mergers, as shown in Figure~\ref{fig: LLS-relation}. The median mass of these clusters is approximately $M_{\rm 500c}\sim8\times10^{14}~\MSUN$, highlighting the significance of simulating truly massive clusters to reproduce these giant radio relics. However, earlier studies have suggested that the size-luminosity relation of observed radio relics is strongly dependent on the sensitivity of the survey. Our TNG-Cluster radio relics indicate that there are likely many faint systems that have not yet been detected. 

    \item The luminosity of the brightest radio relics at fixed cluster mass aligns with the observed mass-luminosity relations (Figure~\ref{fig: MassLumRelation}). After adjusting for the Malmquist bias, the brightness of the relics increases in proportion to the mass of the cluster, as $P\propto M^{2}$, which is in line with theoretical expectations. We also noted that the normalization of the mass-luminosity relation increases by approximately a factor of five from $z=0$ to $z=1$, which we interpret as due to larger magnetic field strengths in the outer regions of clusters at higher redshifts.

    \item The TNG-Cluster radio relics have varying morphologies and physical properties, with significant variations between different systems. These differences are due to factors including: the mass of the merging clusters, the phase of the relic after formation, and the ratio of the cluster mergers. In simulations, radio relics that are brighter than the average mass-luminosity relation are typically found near the center of the cluster during major mergers (Figures~\ref{fig: relic_dist} and \ref{fig: massratio}). Multiple collisions can also increase radio luminosity while maintaining typical arc-shaped morphology (Figure~\ref{fig: mult_merger}).
    
    \item Within the TNG-Cluster simulation, we find naturally produced inverted radio relics. We have identified several examples of these relics, as shown in Figure~\ref{fig: inverted}. In particular, in all cases of inverted relics, there is evidence of a third subgroup or subcluster falling inward. We interpret this inverted morphology as a result of compression before the merger, which causes the post-merger shocks to become warped.
    
    \item Radio relics can be mistakenly identified as radio halos when observed along the collision axis, as shown in Figure~\ref{fig: radiohalo}. To avoid this, strong radio surface brightness discontinuities and weak spatial correlation between radio and X-ray surface brightness can help distinguish face-on radio relics from radio halos. 

\end{itemize}

Our work showcases the scientific capabilities of the TNG-Cluster for both the theory and observability of diffuse synchrotron emission. We adopt a fixed fraction of dissipated energy to generate nonthermal plasma, which is significantly greater than expected from DSA, particularly in the weak shock regime. In the future, we can develop more complex models to determine acceleration efficiency, and these model variations will alter the luminosity of radio relics. Future models will also incorporate the Lagrangian tracers of the TNG-Cluster, enabling more direct comparisons with observed systems through mock observations on the evolution of the radio spectral index and the polarization.

Our results can be compared with observational surveys to constrain the underlying acceleration model. Looking ahead, our catalog of diffuse radio emission from the TNG-Cluster will be a crucial reference point in interpreting the diffuse synchrotron characteristics discovered in upcoming radio surveys with ngVLA, SKA, and similar instruments.

\begin{acknowledgements}
M. J. Jee acknowledges support for the current research from the National Research Foundation (NRF) of Korea under the programs 2022R1A2C1003130 and RS-2023-00219959.
D. Nelson acknowledges funding from the Deutsche Forschungsgemeinschaft (DFG) through an Emmy Noether Research Group (grant number NE 2441/1-1).
JAZ is funded by the Chandra X-ray Center, operated by the Smithsonian Astrophysical Observatory for and on behalf of NASA under contract NAS8-03060.
D. Nagai is supported by NSF (AST-2206055 \& 2307280) and NASA (80NSSC22K0821 \& TM3-24007X) grants. This work is also co-funded by the European Union (ERC, COSMIC-KEY, 101087822, PI: Pillepich).

The TNG-Cluster simulation suite has been executed on several machines: with compute time awarded under the TNG-Cluster project on the HoreKa supercomputer, funded by the Ministry of Science, Research and the Arts Baden-Württemberg and by the Federal Ministry of Education and Research; the bwForCluster Helix supercomputer, supported by the state of Baden-Württemberg through bwHPC and the German Research Foundation (DFG) through grant INST 35/1597-1 FUGG; the Vera cluster of the Max Planck Institute for Astronomy (MPIA), as well as the Cobra and Raven clusters, all three operated by the Max Planck Computational Data Facility (MPCDF); and the BinAC cluster, supported by the High Performance and Cloud Computing Group at the Zentrum für Datenverarbeitung of the University of Tübingen, the state of Baden-Württemberg through bwHPC and the German Research Foundation (DFG) through grant no INST 37/935-1 FUGG. 

This analysis was partially carried out on the VERA supercomputer of the Max Planck Institute for Astronomy (MPIA), operated by the Max Planck Computational Data Facility (MPCDF).

%\section*{Data Availability}
\\
The original IllustrisTNG simulations are publicly available and accessible at \url{www.tng-project.org/data}, as described in \cite{nelson2019a}. The TNG-Cluster simulation will also be made public on the same science platform in the near future. This paper releases the images and data of the $\sim$300 radio relics identified at $z=0-1$: see \url{https://www.tng-project.org/explore/}. Additional data directly related to this publication are available on request from the corresponding authors.
\end{acknowledgements}

\bibliographystyle{aa}
\bibliography{reference}

\begin{appendix}
\section{Resolution tests}
\label{sec: app-res}
\begin{figure}
 \centering
 \includegraphics[width=\columnwidth]{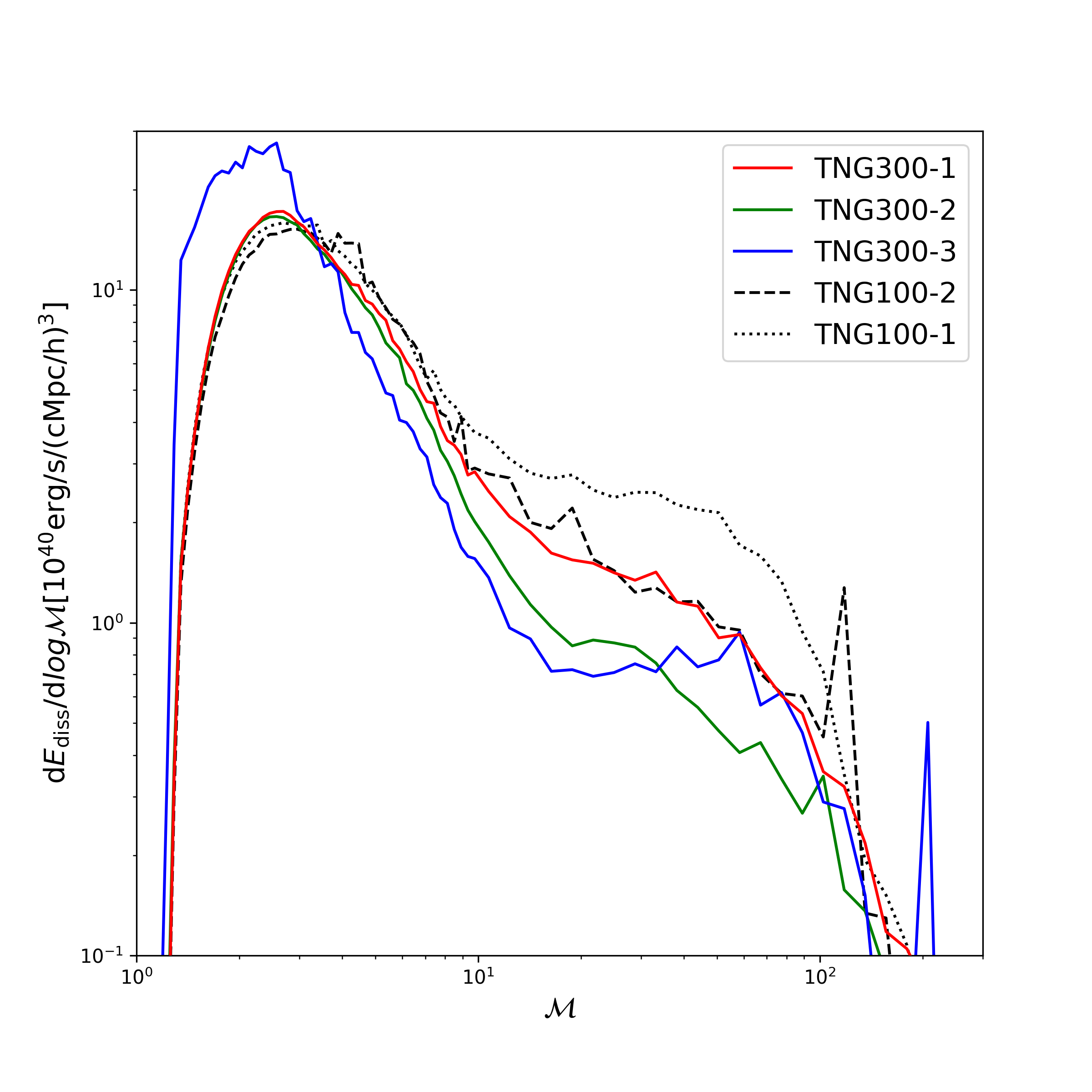}
 \caption{Volume-averaged energy dissipation rate in shocks simulated in the various runs of the IllustrisTNG. The TNG-Cluster simulation has the same resolution as the TNG300-1 and TNG100-2 runs. 
 The energy dissipated by weak shocks ($\mathcal{M}<10$) converges in the simulation runs if the resolution is better than that of TNG300-2, while the energy of the strong shocks ($\mathcal{M}>10$) increases with increasing resolution. 
 }
 \label{fig:shock_convergence}
\end{figure}

\begin{figure}
    \centering
 \includegraphics[width=\columnwidth]{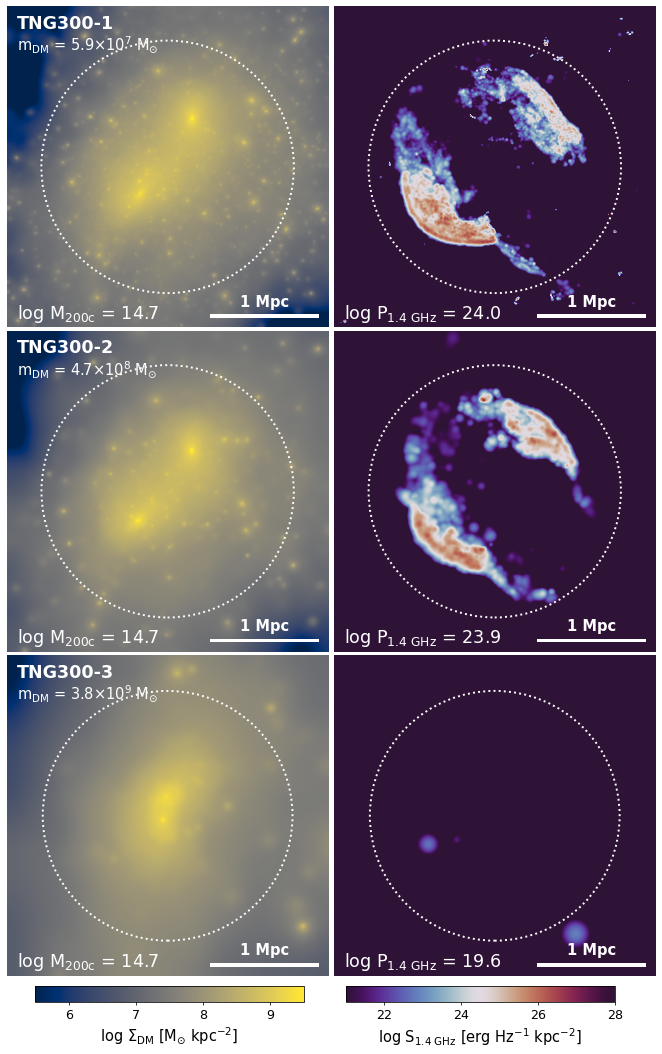}
 \caption{Projection map of mass (left) and radio surface brightness (right column) of the same cluster merger in the TNG300-1 (top), TNG300-2 (middle), and TNG300-3 run (bottom row). The figure is centered on the center of mass. The dotted circle marks the $R_{\rm 500c}$. The annotated $M_{\rm 200c}$ and $P_{\rm 1.4~GHz}$ have the unit of $\rm M_{\odot}$ and $\rm W~Hz^{-1}$, respectively. The radio relic morphology is consistent between the TNG300-2 and TNG300-1 runs, whereas the radio relics are not resolved in the TNG300-3 run.}
 \label{fig:relic_convergence}
\end{figure}

We test the dependence of our results on numerical resolution by using the IllustrisTNG series.
We remind the readers that the TNG-Cluster employs the same TNG galaxy formation model and has a mass resolution identical to the TNG300-1 and TNG100-2 runs. Therefore, we can use the TNG runs \citep[see e.g., summary tables in][]{nelson2019a} to quantify the convergence of resolution of our findings, at least in lower-density environments and for fewer and less massive systems.

Figure~\ref{fig:shock_convergence} presents the distribution of the energy dissipation rate of the cosmological shockwaves in the unit comoving volumes as a function of the Mach number.
Weak shocks with $\mathcal{M}<10$ dominate the dissipated-energy budget by an order of magnitude compared to stronger shocks with $\mathcal{M}>10$.
According to the dissipation rate \citep[e.g.,][]{2007ApJ...669..729K}, the kinetic energy flux of the weak shocks is two orders of magnitude higher than that of the strong shocks, which is consistent with the findings of previous studies \citep[e.g.,][]{2003ApJ...593..599R,2008ApJ...689.1063S,2009MNRAS.395.1333V,2015MNRAS.446.3992S}.

According to this comparison, the energy dissipation rate of the low-$\mathcal{M}$ shockwaves is reasonably well converged at the TNG300-1 resolution, that is, in TNG-Cluster. In fact, the distribution of the dissipated energy rate in weak shocks is consistent across all these runs of the TNG series except for TNG300-3, which has the worst resolution. 
The shock strengths with the maximum energy dissipation rates are slightly lower in TNG300-1 than in TNG100-2, likely due to different box sizes.

On the other hand, the energy dissipated by $\mathcal{M}>10$ shocks increases by a factor of $\mytilde2$ for every improvement in the numerical resolution, namely from TNG300-2 to TNG300-1, and from TNG300-1 to TNG100-1. 
This implies that the properties of strong shocks ($\mathcal{M}>10$) may not be converged at the resolution of the TNG-Cluster. However, studies have found that these shockwaves are dominated mainly by accretion shocks \citep[e.g.,][]{2003ApJ...593..599R}, which has negligible contribution to the diffuse radio features (see Section \ref{sec: merger_lib}).  

We further check the impact of resolution on the radio relic morphology and properties.
In Figure~\ref{fig:relic_convergence}, we show the mass and radio maps of a double radio relic system in the TNG300 series, from the higher-resolution TNG300-1 to TNG300-3 from top to bottom. 
The position of the radio relics and overall morphology are consistent between the TNG300-2 and TNG300-1 runs. 
TNG300-3 presents an entirely different morphology in both dark matter surface density and radio surface brightness maps, which we suspect is the result of the different phases of cluster merger by the sparser timesteps of TNG300-3.  

\begin{figure}
    \centering
 \includegraphics[width=\columnwidth]{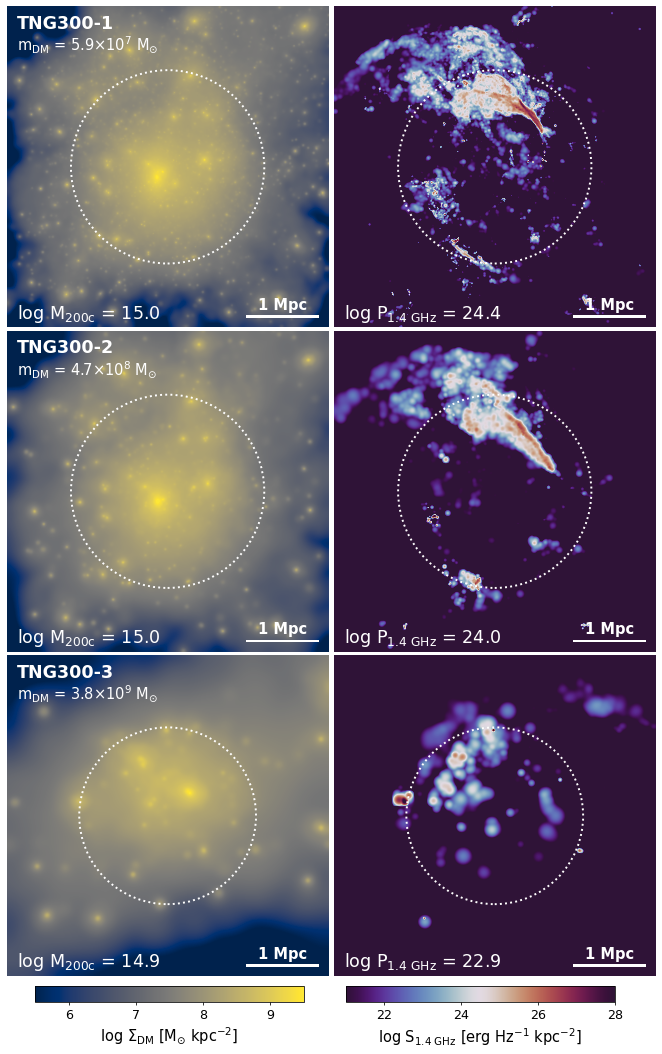}
 \caption{Same as Figure \ref{fig:relic_convergence} but of the single radio relic system.}
 \label{fig:relic_convergence2}
\end{figure}

The physical properties of the radio relics appear reliable at the resolution of TNG300-1, giving us confidence in our results with the TNG-Cluster. In particular, we extract all the radio relic regions using the method described in Section \ref{sec: extract} by grouping the radio structures with a larger bin size for the lower-resolution simulation runs: 40 and $80\rm~kpc$ for the TNG300-2 and TNG300-3 runs, respectively. 
The shock strength ($\mathcal{M}\sim3$) of the relic cells is converged at the resolution of TNG300-2 in all identified radio relics. 
On the other hand, the radio luminosity may vary with the resolution. 
The total radio luminosity of double radio relics is, on average, consistent between TNG300-1 and TNG300-2 ($P_{\rm 1.4~GHz}\sim10^{24}\rm~W~Hz^{-1}$). However, the luminosity of the individual radio relics may change across runs. The luminosity of the brightest radio relic in the TNG300 series increases by $\mytilde60\%$ from TNG300-2 to TNG300-1, whereas that of the double radio relic pair of Figure~\ref{fig:relic_convergence} decreases by a factor of $\mytilde3.5$. 
We find a similar resolution dependence in the single radio relic system of Figure~\ref{fig:relic_convergence2}, where the position, linear extension, and shock strength ($\mathcal{M}\sim4$) are consistent between TNG300-1 and TNG300-2, while the luminosity increased by a factor of $\mytilde2.8$ with resolution improvement.

Based on these resolution tests, we conclude that the overall properties of the radio relics, such as their position from the cluster, morphology, and shock strength, converge at the resolution of the TNG-Cluster simulation.
Radio luminosity varies between TNG300-1 and TNG300-2, so caution is required when interpreting radio relic luminosity. 
Nevertheless, we considered the luminosity statistically consistent since the total energy budget from the cosmological shockwaves converged in TNG300-1. 

\section{Table of observed radio relics}
\label{sec: app-obstab}
Table~\ref{tab:obsrelics} summarizes the properties of all radio relics observed so far and their reference. We have referred to a few of them in the main body of the manuscript.

We quote $M_{\rm 500c}$ of the cluster based on SZ observations when available. In other cases, we derive the X-ray luminosity assuming the mass-luminosity relation of \citet{2009A&A...498..361P} or interpolate the weak-lensing based mass, assuming an NFW profile \citep{1996ApJ...462..563N} and the \citet{2008MNRAS.390L..64D} M-c relation. 
We also include radio relics that are detected only at low-frequency radio wavelengths. 
As we do not have information on the spectral index, we derive the $1.4\rm~GHz$ radio luminosity assuming the spectrum follows $S\propto\nu^{-1.5}$, except Abell 2108 \citep[$\propto\nu^{-2}$,][]{2022ApJ...925...91S}. 

\begin{table*}
    \caption{Properties of the observed radio relics.}
    \centering
    \scriptsize
    \begin{tabular}{ccccccc|ccccccc}
           \hline
           \hline
           Name &  $z$ & $M_{\rm 500c}$ & Pos. &$P_{\rm 1.4 GHz}$ & $LLS$ & Ref.  &  Name &  $z$ & $M_{\rm 500c}$ & Pos. &$P_{\rm 1.4 GHz}$ & $LLS$ & Ref. \\  
           &  & [$10^{14}~\MSUN$] & & $[10^{24}\rm~W~Hz^{-1}]$  & $[\rm Mpc]$  & & &  & [$10^{14}~\MSUN$] & & $[10^{24}\rm~W~Hz^{-1}]$  & $[\rm Mpc]$  & \\
           \hline
           Abell 1656$^{\mathcal{I}}$ & 0.023 & 7.17 & &0.31 & 0.85 & \textbf{1} & PSZ G087$^*$ & 0.222 & 3.83 & & 0.03 & 1.0 & \textbf{15} \\ 
           PSZ2G146$^{*,\mathcal{I}}$ & 0.03 & 1.8 & &0.02 & 0.37 & \textbf{2} & SPT2023-5535$^{\mathcal{I}}$ & 0.23 & 6.49 &  & 3.4& 0.5 & \textbf{23,24} \\ 
           Abell 168  & 0.045 & 1.24 & & 0.14 & 0.8  &\textbf{3} & Abell 746  & 0.232 & 5.34 & W & 4.25 & 1.8  & \textbf{25}  \\ 
           Abell 3376 & 0.046 & 3.64 & E  & 0.54 & 1.6  & \textbf{4} &  &   &   & N & 0.18 & 0.38  &  \textbf{25}  \\
                       &       &      & W  & 0.40 & 0.96 & \textbf{4}  &  &   &   & E & 0.19 & 0.95 &  \textbf{25}   \\
            Abell 3667 & 0.056 & 5.77 & NW &  15.1 & 2.3 & \textbf{5} & PSZ2 G181$^*$ & 0.240 & 4.23 & S & 0.48 & 1.66  & \textbf{15} \\
                               &  & & SE & 2.6 & 1.6 & \textbf{5} & &   &  & N & 0.20 & 1.39  &  \textbf{15}  \\
            Abell 3266$^{\mathcal{I}}$ & 0.059 & 6.64 & & 0.24 & 0.58 & \textbf{6} &  RXCJ1314 & 0.244 & 6.15 & W & 3.90 & 0.91 & \textbf{13} \\
            CIZA J0649 & 0.064 & 3.3 &  & 1.2 & 0.80 & \textbf{7,8} & &  & & E & 1.68 & 1.13  &   \textbf{13}  \\
            ClG 0217+70 & 0.066 & 10.6 & SE & 0.99 & 3.5  & \textbf{9} & MCXC J0929$^*$ & 0.247 & 3.9 & SW & 0.3 & 0.30  & \textbf{17}  \\
                                & &    & SE & 1.27 & 5.1 &  \textbf{9} & &  &   & SE & 0.2 & 0.35  & \textbf{17}  \\
                                &  &    & W & 1.85 & 2.3 & \textbf{9} & Abell 521 & 0.247 & 7.26 & SE & 3.09 & 0.93 & \textbf{26} \\
            RXC J1053 & 0.070 & 1.1 & &0.2 & 0.6 & \textbf{7,8} & ZwCl2341 & 0.27 & 5.15 & S & 4.09 & 1.23 & \textbf{13} \\
            Abell 1904$^*$ & 0.070 & 1.83 & N & 0.32 & 0.23  & \textbf{10} & &  & & N & 1.82 & 0.57 & \textbf{13} \\
                            &   &  &  S  & 0.005 & 0.17  &  \textbf{10} & Abell 1758S$^*$ & 0.280 & 8.22 & & 0.22 & 0.54 & \textbf{15}  \\
                            &   &  &  NE  & 0.01 & 0.9  &  \textbf{10}  & SPT2023-5627 & 0.284 & 5.74 & SE & 1.40 & 0.73 & \textbf{27}  \\
            Abell 2061 & 0.078 & 3.59 & & 0.45 & 0.68 & \textbf{7} &   &  & & NW & 0.91 & 0.86 & \textbf{27}  \\
            Abell 2255 & 0.081 & 5.38 & & 0.18 & 0.70 & \textbf{7}  & Abell 959$^*$ & 0.289 & 5.08 & & 0.22 & 0.81 &  \textbf{15} \\
            Abell 2249 & 0.084 & 3.73 & & 0.41 & 1.3  & \textbf{11} &1E 0657-55.8 & 0.296 & 11.4 & E & 43.5 & 0.99 &  \textbf{28} \\
            Abell 2018$^*$ & 0.088 & 2.51 & & 0.04 & 1.1  &  \textbf{10}&  &   &   & NNW  & 0.6 & 0.90 & \textbf{28}\\
            Abell 2108$^*$ & 0.092 & 1.8 & &0.01 & 0.2  &  \textbf{12}& Abell 781 & 0.295 & 6.13 & & 4.47 & 0.44 & \textbf{1} \\
            Abell 3365 & 0.093 & 1.7 & E & 0.74 & 0.55  & \textbf{13,14}& PSZ2 G198$^*$ & 0.299 & 5.5 & & 0.51 & 1.49 &  \textbf{15} \\
                               &  &  & W & 0.09 & 0.29  &  \textbf{13,14}& PSZ1G097 & 0.3 & 4.7 & S & 3.12 & 1.42 & \textbf{13}\\
            Abell 523 & 0.10 & 1.7 & & 1.7 & 1.35 & \textbf{7,8} &   &  &  & N & 1.52 & 0.88 & \textbf{13} \\       
            ZwCl0008 & 0.103 & 3.30 & E & 1.54 & 1.41 & \textbf{13}& Abell 1300 & 0.3072 & 8.97 & & 5.75 & 0.70 &  \textbf{1}  \\
                                     &  & & W & 0.30 & 0.30 &\textbf{13} & Abell 2744 & 0.308 & 9.84 & NE & 4.37 & 1.5 &  \textbf{29} \\
            Abell 1925$^*$ & 0.105 & 2.81 & N & 0.12 & 1.71 & \textbf{15} & &  & & SE & 0.96 & 1.15 & \textbf{29}  \\
                           &  &  & S & 0.08 & 1.09 &  \textbf{15} & MACS J0258  & 0.322 & ~ &  & ~ & 0.39 & \textbf{16}  \\
            Abell 2034 & 0.113 & 5.85 & & 0.89 & 0.22  &\textbf{7} & Abell 1943$^*$ & 0.336 & 8.14 & & 0.03 & 0.37 & \textbf{17}  \\
            Abell 3186 & 0.127 & 6.44 &  & 2.5 & 2.0 & \textbf{16} & PSZ2 G121$^*$ & 0.344 & 5.69 & & 0.21 & 0.54 & \textbf{15} \\
            WHL J1013$^*$ & 0.146 & 2.49 & & 0.01 & 0.31 & \textbf{17} & PSZ2 G096$^*$ & 0.350 & 5.39 & & 0.60 &1.48 & \textbf{15} \\
            PSZ2 G278 & 0.158 &  3.6 & N & 1 & 0.66 & \textbf{18} & PSZRX G095$^*$ & 0.362 & 3.34 & & 0.8 & 1.30 & \textbf{17}\\
                               &       &      & S & 0.7 & 1.64 & \textbf{18} & PSZ2 G092$^*$  & 0.362 & 6.31 & & 0.32 & 1.53 & \textbf{15} \\
            Abell 1240 & 0.159 & 3.71 & S & 0.73 & 1.28  & \textbf{13} & MACS J1752 & 0.366 & 6.96 & NE & 33.1 & 1.19 & \textbf{13} \\
                               &   &  & N & 0.43 & 0.87  & \textbf{13} &  &  & & SW & 15.0 & 0.69 &  \textbf{13}  \\
            Abell 3411 & 0.169 & 6.59 &  & 5.0 & 1.9  &\textbf{19} & PSZ2 G114$^*$ & 0.371 & 7.58 & N  & 3.40 & 1.17 & \textbf{15} \\
            PSZ2 G109$^*$ & 0.173 & 3.26 & & 0.50 & 1.17 & \textbf{15}  &  &  &  & S  & 0.83 & 1.48 &  \textbf{15} \\
            PSZRX G182$^*$ & 0.175 & 2.4 & & 0.3 & 1.00  & \textbf{17} & ZwCl1447 & 0.376 & 3.4 & SW & 1.11 & 1.2 & \textbf{30} \\
            Abell 2345 & 0.177 & 5.92 & W & 2.85 & 0.83 & \textbf{13} & &  & & NE &0.51 & 0.3 & \textbf{30} \\
                                &  &  & E & 2.66 & 1.57 & \textbf{13}  & PSZ2 G187$^*$ & 0.378 & 6.84 & & 0.27 & 0.72 & \textbf{15} \\
            Abell 1612 & 0.179 & 4.42 & & 7.9 & 0.78  & \textbf{7} & PSZ1 G287 & 0.39 & 13.89 & NW & 23.3 & 2.48  &  \textbf{13} \\
            Abell 1697$^{*,\mathcal{I}}$ & 0.183 & 4.34 & & 0.38 & 0.7  & \textbf{10}  &  &  &   & SE & 9.71 & 1.58 & \textbf{13}  \\
            WHL J1721$^*$ & 0.184 & 2.29 & & 0.08 & 0.60 &  \textbf{17} & WHL J1305$^*$ & 0.396 & 3.29 & & 0.06 & 1.40 &\textbf{17}\\
            Abell 1889$^*$ & 0.185 & 2.58 & NW & 0.50 & 1.65 &  \textbf{17} &PSZ2 G117$^*$ & 0.396 & 7.61 & & 0.32 & 0.56 & \textbf{15} \\
                                    &  &  & SE & 0.40 & 1.65 &  \textbf{17}  & MCXC J0943$^*$ & 0.406 & 4.75 & & 0.4 & 0.53  & \textbf{17}  \\
            PSZ2 G145$^*$ & 0.190 & 4.25 & & 0.04 & 0.77 & \textbf{15} & PSZ2 G206$^*$ & 0.447 & 7.39 & S & 0.25 & 0.66 & \textbf{15} \\
            CIZAJ2243 & 0.192 & 16.2 & N & 15.0  & 2.00  & \textbf{13,20} & &  &   & N & 0.18 & 0.66 & \textbf{15}  \\
                             &      &       & S & 1.95 & 1.65  & \textbf{13,20} & PSZ2 G191$^*$ & 0.488 & 5.55 & & 2.49 & 0.57  & \textbf{15}  \\
            Abell 115 & 0.197 & 7.65 & & 2.7 & 2.5  & \textbf{21} & MACS J1149  & 0.544 & 8.55 & W & 6.03 & 0.70 &   \textbf{13} \\
            Abell 2163 & 0.203 & 16.1 & & 2.09 & 0.48  & \textbf{1} &  &  &   & E & 5.33 & 0.78 & \textbf{13} \\ 
            1RXS J0603 & 0.22 & 10.76 & N & 65.0 & 1.9 & \textbf{22} & PSZ2 G069 & 0.762 & 5.69 & & 3.75 & 1.49 &  \textbf{15} \\
                       &  & & E & 2.5  & 1.1 & \textbf{22} & PSZ2 G092 & 0.822 & 7.4 & & 5.25 & 0.9 & \textbf{31} \\
                       &  & & S & 1.4  & 0.25 &  \textbf{22} & ACT J0102 & 0.87 & 8.80 & NW & 29.6 & 0.93  & \textbf{13} \\
            PLCK G201 & 0.22 & 2.7 & E & 3.9 & 1.19 & \textbf{16} &  &  &  & SE & 4.48 & 0.46  & \textbf{13}  \\
            \hline
    \end{tabular}
    \tablefoot{Column 1: cluster name. Column 2: cluster redshift. Column 3: radio luminosity at $1.4\rm~GHz$. We derive the luminosity assuming the radio spectrum of $S\propto\nu^{-1.5}$ for the radio relics only detected at a low frequency. These radio relics are annotated with $^*$. The inverted radio relics are marked with $^{\mathcal{I}}$. Column 4: largest linear size (LLS). Column 5: reference for the cluster radio luminosity.} %\tablebib{\textbf{1}. \citet{2012A&ARv..20...54F};  \textbf{2}. \citet{2021ApJ...914L..29B}; \textbf{3}. \citet{2018MNRAS.477..957D}; \textbf{4}. \citet{2023PASJ...75S..97C}; \textbf{5}. \citet{2022A&A...659A.146D}; \textbf{6}. \citet{2022MNRAS.515.1871R}; \textbf{7}. \citet{2011A&A...533A..35V}; \textbf{8}. \citet{2009A&A...498..361P}; \textbf{9}. \citet{2021A&A...656A.154H}; \textbf{10}. \citet{2021A&A...651A.115V}; \textbf{11}. \citet{2020MNRAS.496L..48L}; \textbf{12}. \citet{2022ApJ...925...91S}; \textbf{13}. \citet{2014MNRAS.444.3130D}; \textbf{14}. \citet{2019MNRAS.483..540L}; \textbf{15}. \citet{2022A&A...660A..78B}; \textbf{16}. \citet{2022A&A...657A..56K}; \textbf{17}. \citet{2022A&A...665A..60H}; \textbf{18}. \citet{2023arXiv230411784K}; \textbf{19}. \citet{2013ApJ...769..101V}; \textbf{20}. \citet{2015ApJ...802...46J}; \textbf{21}. \citet{2001A&A...376..803G}; \textbf{22}. \citet{2016ApJ...818..204V}; \textbf{23}. \citet{2020ApJ...900..127H}; \textbf{24}. \citet{2019ApJ...871...50B}; \textbf{25}. \citet{2023arXiv230901716R}; \textbf{26}. \citet{2006NewA...11..437G}; \textbf{27}. \citet{2021PASA...38....5D}; \textbf{28}. \citet{2023MNRAS.518.4595S}; \textbf{29}. \citet{2017ApJ...845...81P}; \textbf{30}. \citet{2022ApJ...924...18L}; \textbf{31}. \citet{2023A&A...675A..51D}}
    \tablebib{(1)~\citet{2012A&ARv..20...54F}; (2) \citet{2021ApJ...914L..29B}; (3) \citet{2018MNRAS.477..957D}; (4) \citet{2023PASJ...75S..97C}; (5) \citet{2022A&A...659A.146D}; (6) \citet{2022MNRAS.515.1871R}; (7) \citet{2011A&A...533A..35V}; (8) \citet{2009A&A...498..361P}; (9) \citet{2021A&A...656A.154H}; (10) \citet{2021A&A...651A.115V}; (11) \citet{2020MNRAS.496L..48L}; (12) \citet{2022ApJ...925...91S}; (13) \citet{2014MNRAS.444.3130D}; (14) \citet{2019MNRAS.483..540L}; (15) \citet{2022A&A...660A..78B}; (16) \citet{2022A&A...657A..56K}; (17) \citet{2022A&A...665A..60H}; (18) \citet{2023arXiv230411784K}; (19) \citet{2013ApJ...769..101V}; (20) \citet{2015ApJ...802...46J}; (21) \citet{2001A&A...376..803G}; (22) \citet{2016ApJ...818..204V}; (23) \citet{2020ApJ...900..127H}; (24) \citet{2019ApJ...871...50B}; (25) \citet{2023arXiv230901716R}; (26) \citet{2006NewA...11..437G}; (27) \citet{2021PASA...38....5D}; (28) \citet{2023MNRAS.518.4595S}; (29) \citet{2017ApJ...845...81P}; (30) \citet{2022ApJ...924...18L}; (31) \citet{2023A&A...675A..51D}}
    \label{tab:obsrelics}
\end{table*}

\end{appendix}
\end{document}